\documentclass[superscriptaddress,amsmath,floatfix,aps,prc,twocolumn,nofootinbib]{revtex4-2}
\usepackage{graphicx}
\usepackage{tabularx}
\usepackage{xcolor,colortbl}
\usepackage{slashed}
\usepackage{hyperref}
\usepackage{cleveref}
\usepackage{preamblestyle} % Needs preamblestyle.sty file

\newcommand{\neutrons}{n_{\pm}}
\newcommand{\protons}{p_{\pm}}
\newcommand{\neutron}{n_{+}}
\newcommand{\neutronMinus}{n_{-}}
\newcommand{\proton}{p_{+}}
\newcommand{\protonMinus}{p_{-}}

\newcommand{\nucleon}{N_{+}}
\newcommand{\nucleonMinus}{N_{-}}

\begin{document}

\title{Enhanced Neutrino Cooling from Parity-Doubled Nucleons in Neutron Star Cooling Simulations}

\author{Rodrigo Negreiros \orcidlink{0000-0002-9669-905X}}
\email{rnegreiros@catholic.tech}
\affiliation{Department of Physics, Catholic Institute of Technology, MA, USA}
\affiliation{Department of Physics, Universidade Federal Fluminense, Niteroi, Brazil}
\affiliation{
ICRANet, Piazza della Repubblica 10, I-65122 Pescara, Italy}

\author{Liam Brodie \orcidlink{0000-0001-7708-2073}}
\email{b.liam@wustl.edu}
\affiliation{Department of Physics, Washington University in St.~Louis, St.~Louis, MO 63130, USA}

\author{Jan Steinheimer \orcidlink{0000-0003-2565-7503}}
\email{steinheimer@fias.uni-frankfurt.de}
\affiliation{Frankfurt Institute for Advanced Studies (FIAS),
Ruth-Moufang-Str. 1, D-60438 Frankfurt am Main, Germany}
\affiliation{GSI Helmholtzzentrum für Schwerionenforschung GmbH, Planckstr. 1, 64291 Darmstadt, Germany}

\author{Veronica Dexheimer \orcidlink{0000-0001-5578-2626}}
\email{vdexheim@kent.edu}
\affiliation{Department of Physics, Kent State University, Kent OH 44242 USA}

\author{Robert D. Pisarski \orcidlink{0000-0002-7862-4759}}
\email{pisarski@bnl.gov}
\affiliation{Department of Physics, Brookhaven National Laboratory, Upton, NY 11973}

\date{March 9, 2026} % Add date before submitting to arXiv

\begin{abstract}
Although restoration of chiral symmetry is predicted by quantum chromodynamics to take place at high baryon density, most modeling of neutron star interiors disregards a chiral phase transition. We model neutron star cores with a parity doublet model, which allows for dynamical chiral symmetry restoration and predicts the appearance of the parity partners of nucleons and hyperons at large densities, as well as deconfined quark matter. We study the thermal evolution of neutron stars, focusing for the first time on the impact of Urca processes involving the parity partners in neutron star cooling simulations. We find that Urca processes for the parity partners of the nucleons significantly affect the thermal evolution of massive stars and allow for improved agreement with observed surface temperature and ages.
\end{abstract}

\maketitle

\section{Introduction}
\label{sec:intro}

Neutron stars provide a unique window into the physics of dense matter, as they are the only objects in the Universe where densities of several times nuclear saturation can be accessed at effectively zero temperature. While the dense-matter equation of state (EoS) can be constrained by measurements of neutron star mass, radius, and tidal deformability~\cite{Tan:2021ahl,Tan:2021nat}, there are limits to what can be learned. For example, stars containing a significant amount of deconfined quark matter in their cores can masquerade as purely hadronic stars by exhibiting a mass--radius relationship similar to that predicted for a star made entirely of
hadrons~\cite{Alford:2004pf}. One solution to this problem would be to analyze the gravitational-wave signal from neutron star mergers~\cite{Hammond:2025kki}. However, the post-merger gravitational-wave signal, where matter is expected to be far from equilibrium, has not yet been detected. 

In conjunction with studies of mergers, the thermal evolution of neutron stars (e.g., cooling) can provide information about neutron star interiors, including particle content, interactions, and pairing schemes. This is because the microscopic composition of neutron stars plays a vital role in their cooling, notably affecting thermal conductivity, specific heat capacity, and neutrino emissivity. Importantly, there is substantial data about the thermal evolution of neutron stars (ages and thermal luminosities)~\cite{Potekhin:2020ttj}.

We focus on the thermal evolution of neutron stars using an SU(3) parity doublet model, i.e., considering the entire baryon octet with both positive parity (the nucleons and hyperons) and negative parity (the respective parity partners of the nucleons and hyperons). We analyze the impact of the parity partners on the thermal evolution of intermediate-mass and massive neutron stars ($1.4-2.1\,\text{M}_{\odot}$) with cores where chiral symmetry has been partially restored. Although previous research, such as Refs.~\cite{Dexheimer:2012eu,Mukherjee:2017jzi,Marczenko:2018jui}, explored neutron star cooling within the SU(3) parity doublet model, they did not include neutrino processes involving the parity partners. Recently, Ref.~\cite{Brodie:2025nww} found that the nucleon parity partners allow for rapid neutron star cooling by calculating the neutrino emissivity and thermal evolution, albeit in an isothermal regime. 
Our objective is to quantify how the parity partners affect the thermal evolution of neutron stars by performing a self-consistent cooling simulation of the entire star, accounting for the thermodynamics of the crust, core, and atmosphere~\cite{Page:2004fy,Page:2006ud,Negreiros:2010tf,Negreiros:2011ak,Negreiros:2012aw}.

\section{Methods}
\subsection{Parity Doublet Model}
\label{sec:pdm_model}

At high temperatures and/or densities, QCD (quantum chromodynamics, the fundamental theory that describes quarks and gluons) predicts that chiral symmetry is restored. The spontaneous breaking of chiral symmetry at low temperatures and densities is intrinsically linked to the creation of most of a baryon's mass. At vanishing baryon chemical potential, lattice QCD has shown that the mass of the baryon octet and its parity partners become degenerate, but still finite, as the temperature increases, thereby restoring chiral symmetry while retaining a finite baryon mass~\cite{Aarts:2017rrl}. To incorporate such a realization of chiral symmetry restoration at low temperature and high baryon density, parity doublet models for nucleons were developed and extended~\cite{Detar:1988kn,Zschiesche:2006zj,Steinheimer:2011ea,Sasaki:2017glk,Minamikawa:2020jfj,Marczenko:2022hyt,Fraga:2023wtd,Eser:2024xil,Gao:2026scv}.
While quark deconfinement is still not well understood in the strong-coupling regime of QCD, effective descriptions of deconfinement can be introduced~\cite{Ratti:2005jh,Roessner:2006xn,Dexheimer:2009hi,Steinheimer:2011ea}.

To incorporate features such as the appearance of hyperons, deconfined quarks, and various constraints on the equation of state from low-energy nuclear physics, astrophysics, and heavy-ion collisions, we consider the parity doublet chiral mean field~(PD-CMF) model. The model consists of an SU(3) Lagrangian with a non-linear realization of chiral symmetry for hadronic matter first introduced in Ref.~\cite{Papazoglou:1998vr}, then extended to describe neutron stars in Ref.~\cite{Dexheimer:2008ax}, and quark deconfinement in Ref.~\cite{Dexheimer:2009hi} through the introduction of a Polyakov loop order parameter.

The particular PD-CMF we use \cite{Steinheimer:2011ea,Motornenko:2020yme,Steinheimer:2025hsr} (see Appendix~\ref{App:CMF} for the full Lagrangian and our choice of parameters) also includes a hadron resonance gas of baryons (beyond the octet) and mesons \cite{ParticleDataGroup:2022pth}, which is useful for describing high-temperature physics and is in agreement with lattice QCD results~\cite{PhysRevD.108.014510,Borsanyi:2012cr}.
The Lagrangian ${\cal L}_{{\rm SU}(3)}$ consists of the following parts
\begin{align}
{\cal L}_{\rm{SU(3)}}={\cal L_B}  + U_{\rm scalar} + U_{\rm vector} \,,
\end{align}
where ${\cal L_B}$ describes scalar and vector mean-field interactions among the ground-state octet baryons and their parity partners. $U_{\rm scalar}$ is the potential of the scalar $\sigma$ and $\zeta$ fields, and $U_{\rm vector}$ is the potential of the vector $\omega$, $\rho$, and $\phi$ fields.

The effective masses of the ground-state octet baryons and their parity partners are
\begin{align}
m^*_{b\pm} &= \sqrt{  (g^{(1)}_{\sigma b} \sigma + g^{(1)}_{\zeta b}  \zeta )^2 + (m_0+n_s m_s)^2 } \pm g^{(2)}_{\sigma b} \sigma\,,
\end{align}
where the indices $(1)$ and $(2)$ refer respectively to the two couplings that determine the common mass and mass difference of the parity partners. The coupling constants $g^{(i)}_{ib}$ are determined by vacuum masses and by nuclear matter properties. $m_0$ refers to a bare mass term of the baryons which is not generated by the breaking of chiral symmetry, and $n_s m_s$ is the ${\rm SU}(3)$-breaking mass term that generates an explicit mass corresponding to the number of strange quarks $n_s$ in a baryon. In the present version of the PD-CMF model, different isospin states of baryons have the same mass (e.g., $m^*_{p\pm} = m^*_{n\pm}$) since the mass is only generated by the scalar field and no scalar iso-vector coupling (e.g., to the $\delta$) is included. 
The vector interactions lead to a modification of the effective chemical potentials for the baryons and their parity partners
\begin{equation}
    \mu^*_b=\mu_b-g_{\omega b} \omega-g_{\phi b} \phi-g_{\rho b} \rho \,,
\end{equation}
where the physical chemical potential is $\mu_b$.

Identifying the physical parity partners of the baryon octet from the particle data group list of hadrons \cite{ParticleDataGroup:2022pth} is not straightforward. The uncertainties on the masses of candidate parity partners are large, and in some cases, several candidates exist. In the PD-CMF model we made the following assumptions:
i) the parity partners of the nucleons are the $\mathrm{N_{-}}(1535)$ and ii) for the strange baryons, we assume that the mass splitting between the positive and negative parity states is similar to that for the nucleon. Based on these assumptions, we chose the closest states from the particle data group. While
reasonable phenomenologically, the detailed analysis of parity doubled states with
three flavors is involved; see Refs.~\cite{Minamikawa:2020jfj,Minamikawa:2021fln,Minamikawa:2022ckn,Gao:2022klm,Minamikawa:2023eky,Minamikawa:2023ypn,Gao:2024mew},
and Ref.~\cite{Fraga:2023wtd}.
Appendix~\ref{App:CMF} lists the vacuum masses of all parity partners within our model.

The quark degrees of freedom are introduced in a way similar to the Polyakov-loop-extended Nambu Jona-Lasinio (PNJL) model~\cite{Fukushima:2003fw}, where their thermal contribution is directly coupled to the Polyakov loop order parameter $\Phi$ \cite{Motornenko:2019arp}. Unlike the standard PNJL approach, only the single (deconfined) quark contribution is added to the thermodynamic potential. This is done to avoid double-counting of the confined 3-quark state and to obtain the proper asymptotic limit of a free quark gas at very large chemical potentials. The quark contribution to the thermodynamic potential per unit volume is
\begin{equation}
	\frac{\Omega}{V}=-T \sum_q\frac{d_q}{N_c}\int\!\frac{d^3k}{(2 \pi)^3}\ln\left(1+ \Phi\,e^{{-\left(E_q^*-\mu^*_q\right)/T}}\right)\,,
	\label{eq:q}
\end{equation}
with index $q$ running through $u,d,s$ flavors, $N_c$ the number of colors, and $d_{q}$ the color and spin degeneracy factor.

The temperatures we consider in this paper \mbox{$T<10^{7}$ K} or \mbox{$T<10^{-3}$ MeV} are not sufficient to modify the EoS in a meaningful way. Therefore, for the EoS only, we assume $T=0$. 
In this case, the Polyakov loop $\Phi$ is not well constrained since the exponential term in~\Cref{eq:q} either vanishes or is infinite for $T\rightarrow 0$. 
This simplifies the quark contribution to that of a Fermi gas of quarks with effective masses $m^*_{q}$ as long as $\Phi$ is non-vanishing. For simplicity, we set $\Phi=1$.
The dynamical quark masses $m_{q}^*$ are determined by the $\sigma$ and $\zeta$ fields, along with a bare mass term $m_{0q}$, which can be understood as the contribution of the gluon condensate to the quark mass
\begin{align}
m_{u}^* & =-g_{u\sigma}\sigma+\delta m_{u} + m_{0u}\,,\nonumber\\
m_{d}^* & =-g_{d\sigma}\sigma+\delta m_{d} + m_{0d}\,,\nonumber\\
m_{s}^* & =-g_{s\zeta}\zeta+ \delta m_s + m_{0s}\,.
\end{align}
The vector couplings of the quarks are set to zero to obtain the correct asymptotic limit of a free gas of quarks. The introduction of a bare mass term for the quarks $m_{0q}$ explicitly breaks chiral symmetry. This means that when we refer to chiral symmetry restoration in PD-CMF, we imply that chiral symmetry can only be approximately restored.

The suppression of hadrons at high baryon densities is obtained by their excluded-volume hard-core interactions~\cite{Rischke:1991ke,Steinheimer:2011ea}.
The volume term modifies the effective chemical potential,
\begin{equation}
    \mu^*_b \to \mu^*_b - v_b\,P\,,
\end{equation}
for each baryon species $b$, including the parity partners. The total pressure of the system is $P$, and $v_b=0.4\,\rm{fm}^{3}$ is the excluded volume parameter for the baryons (the quarks are always assumed to be point-like).

The PD-CMF parameter set we use (see Appendix~\ref{App:CMF}) provides a good description for isospin-symmetric nuclear matter saturation properties: saturation density $n_b=0.16$~fm$^{-3}$, binding energy per nucleon $E_0/B =-16$~MeV, symmetry
energy of $S_0 = 31$~MeV and symmetry
energy slope $L = 53$~MeV, as well as an incompressibility $K_0 = 303$~MeV.

To compare our results to a baseline without parity partners, we can also calculate the thermodynamic properties of the model in a scenario where the parity partners never appear. In this scenario, 
which we refer to as the parity singlet CMF, or PS-CMF, 
chiral symmetry is not restored for the baryons. For neutron star calculations, a gas of free leptons (electrons and muons) is also included to ensure charge neutrality and chemical equilibrium. 

\begin{figure*}[tbp]
\label{fig:pdm_model}
  \centering
    \centering
    \includegraphics[width=0.49\textwidth]{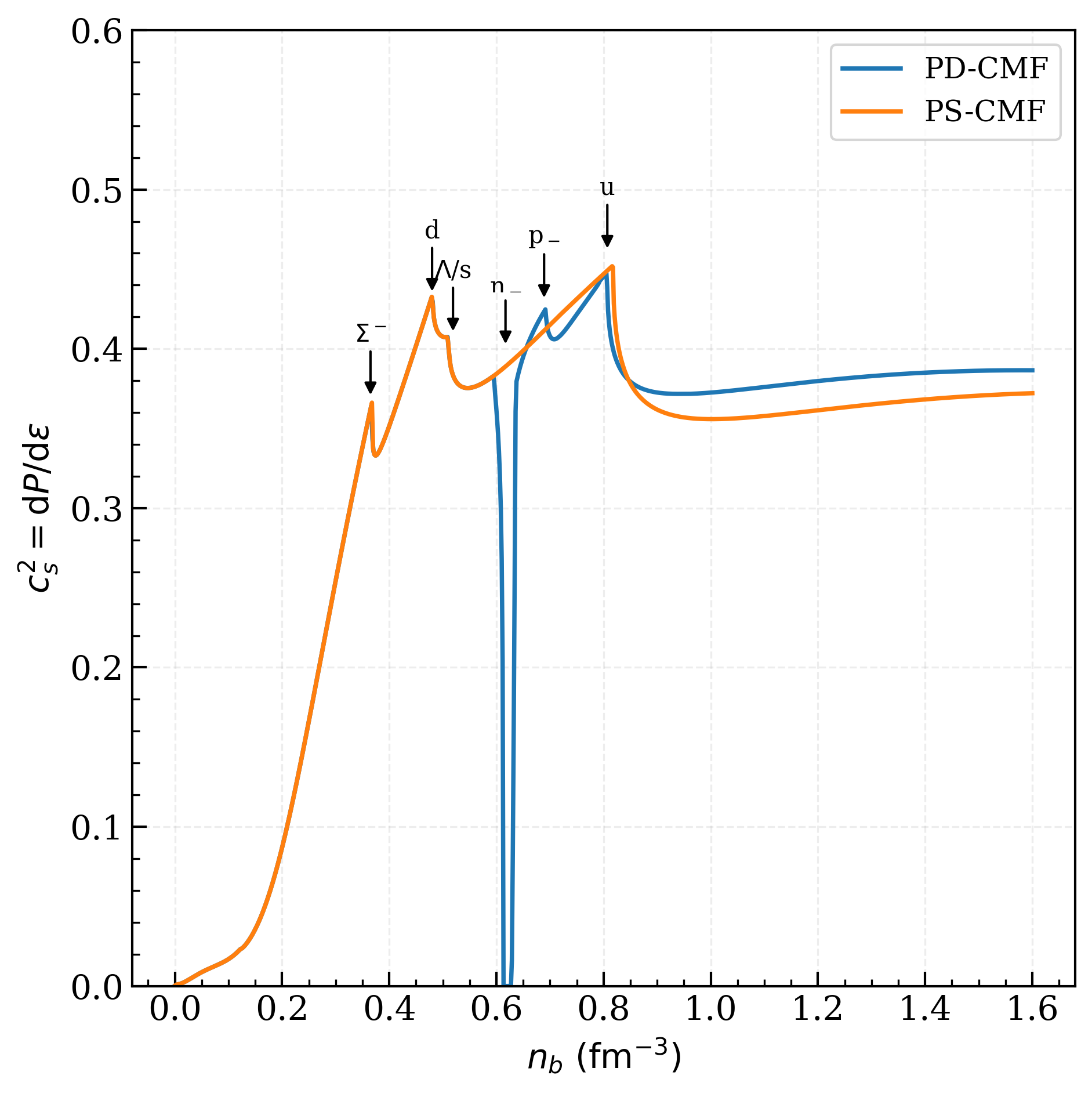} 
    \centering
    \includegraphics[width=0.49\textwidth]{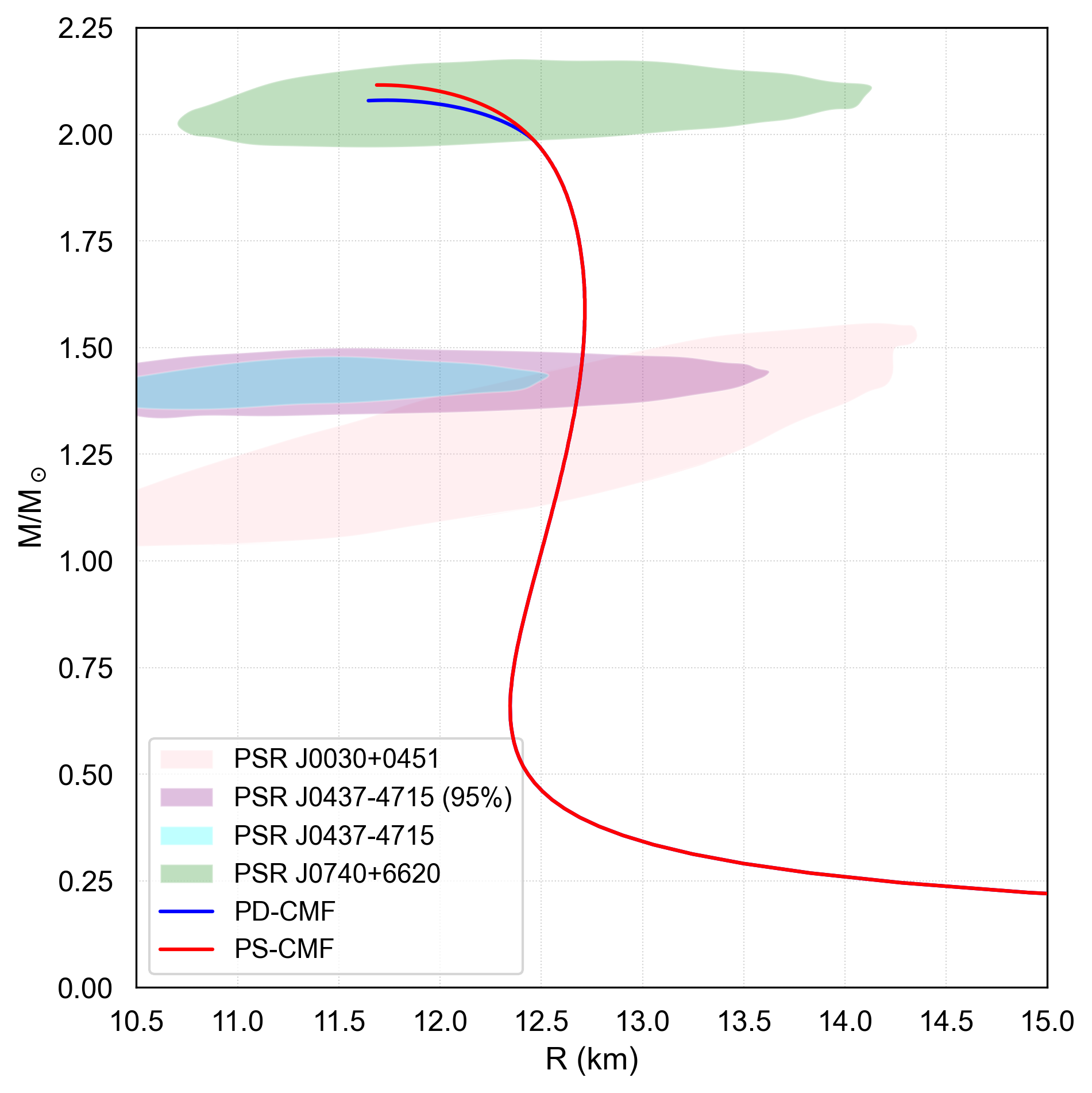} 
  \vspace{4mm} % small vertical gap between rows
    \centering
    \includegraphics[width=0.49\textwidth, height = 0.485\textwidth] {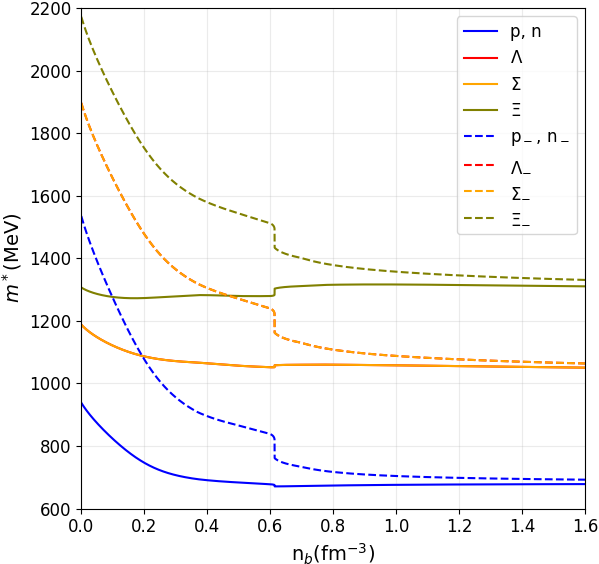} 
    \centering
    \includegraphics[width=0.49\textwidth]{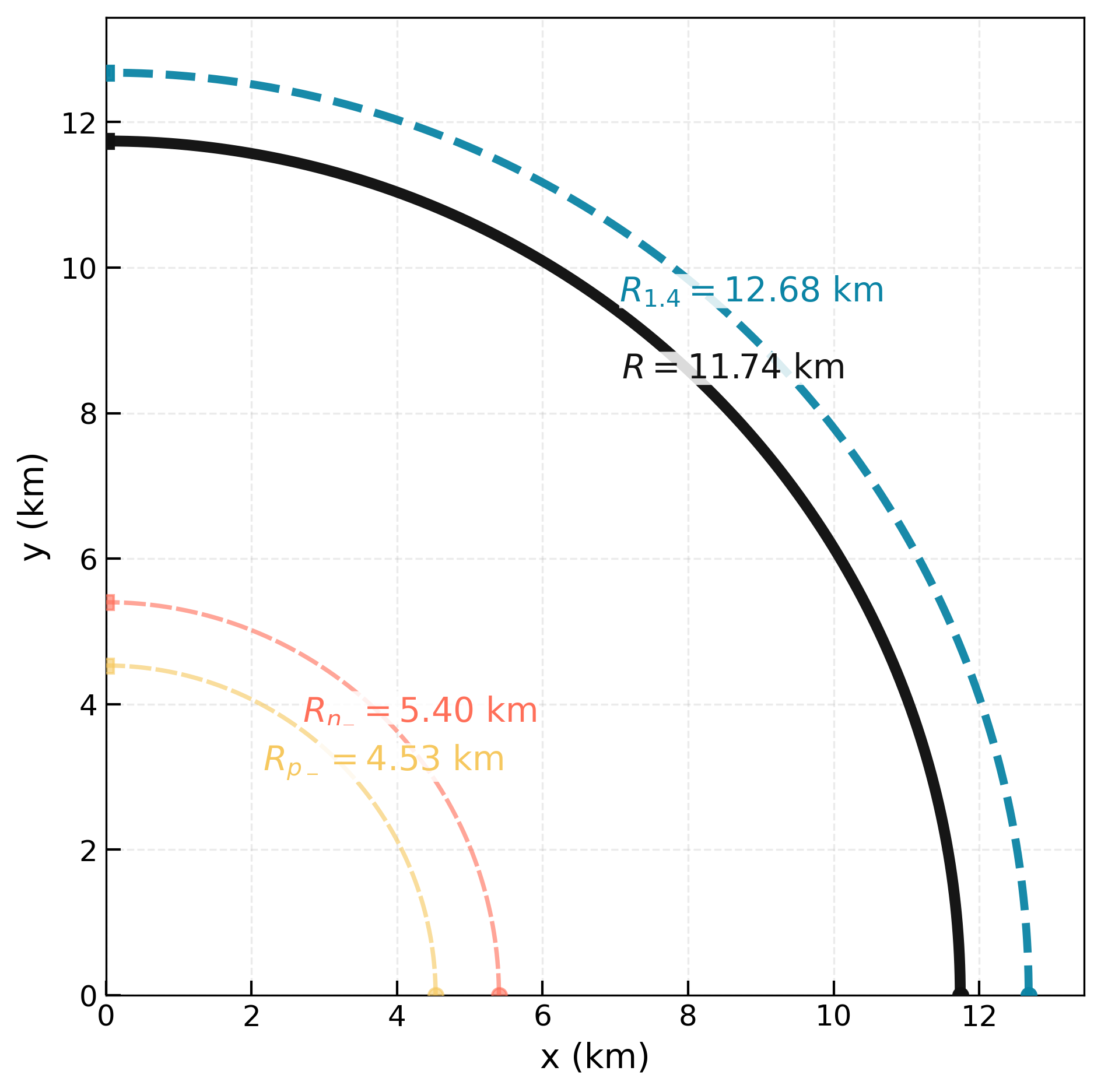}   
  \caption{Properties of the parity doublet chiral mean field (PD-CMF) and parity singlet chiral mean field (PS-CMF) models we consider.\\
  (a) Top left: The speed of sound squared as a function of baryon density for neutron star matter described by the PD-CMF and PS-CMF models, showing the onset density of various particles.\\
  (b) Top right: The mass--radius predictions for the PD-CMF and PS-CMF models. The shaded regions are astrophysical constraints at the $68\,\%$ confidence level.\\
  (c) Bottom left: The particle masses as a function of baryon density in the PD-CMF model.\\ (d) Bottom right: A radial profile of the most massive star predicted by the PD-CMF model showing where the parity partners appear. The dashed curve shows the radius of a $1.4$ M$_\odot$ star and the solid black line the radius of the maximum-mass star.}
  \label{fig:2x2grid}
\end{figure*}

In the top left panel of~\Cref{fig:2x2grid}, we show the
speed of sound for neutron star matter at zero temperature in
both the PS-CMF and PD-CMF models, as a function
of density.
There are cusps in the speed of sound as various degrees of
freedom, including the $\Sigma^-$, $d$ quark, $\Lambda$, and $s$ quark appear.  

The speed of sound in the two models agrees
up to a density of $n_b\,\approx\,0.6\,\mathrm{fm}^{-3}$. At
this density, in the parity doublet model, there is a first-order transition, where the parity partner of the neutron appears, so the speed of sound drops to zero. However,
the transition is only weakly first order,
with a small jump in the energy density
$\Delta \eps/\eps \approx 4 \%$.  Such a weak first-order transition leads to no significant
changes in the mass--radius curve (the top right panel of~\Cref{fig:2x2grid}).
Beyond the first-order transition, there are noticeable but small differences
in the speed of sound between the two models, including a cusp where the parity partner of the proton appears, as illustrated
in the top left panel of~\Cref{fig:2x2grid}.

The density dependence of the effective masses of the different particle species
is shown in the bottom left panel of~\Cref{fig:2x2grid}.
Below the first-order chiral transition ($n_b\approx\,0.6\,\rm{fm}^{-3}$), the difference between the parity
partner masses is large. Above the chiral transition, the difference between the mass of the baryons and their parity partners decreases, but chiral symmetry is not fully restored.  

At low densities, nuclear matter is inhomogeneous, so we use the Baym--Pethick--Sutherland (BPS) crustal EoS~\citep{Baym:1971pw} and attach it to the PD-CMF and PS-CMF models that we use at higher densities. 
We solve the Tolman–Oppenheimer–Volkoff (TOV) equations~\cite{Tolman,OppenheimerVolkoff} for both the PD-CMF and PS-CMF models and show the resulting mass-radius curves in the top right panel of~\Cref{fig:2x2grid}. As the parity partners only appear at moderate baryon densities, their effect on the mass-radius curve is subtle and only affects stars with masses above about $2\,\rm{M}_{\odot}$. Their modification to the thermal evolution of neutron stars may therefore serve as a more obvious signal of the effects of chiral symmetry restoration.

The region where the parity partners of the nucleons are present in the maximum-mass star predicted by the PD-CMF model ($M\approx2.1\,\rm{M}_{\odot}$) is shown in the bottom right panel of~\Cref{fig:2x2grid}. The parity partners of the hyperons appear at densities exceeding the central density of this star. We show the radius of the maximum-mass star as well as the radius of a $1.4\,\rm{M}_\odot$ star.

\subsection{Neutron Star Thermal Evolution}
\label{sec:thermal_evolution}

To model the thermal evolution of a neutron star, we consider the cooling equations \cite{2006NuPhA.777..497P,1999Weber..book,1996NuPhA.605..531S}
\begin{align}
  \frac{ \partial (l e^{2\phi})}{\partial m}& = 
  -\frac{1}{\epsilon \sqrt{1 - 2m/r}} \left( Q\, 
    e^{2\phi} + c_v \frac{\partial (T e^\phi) }{\partial t} \right) \, , 
  \label{eq:coeq1}  \\
  \frac{\partial (T e^\phi)}{\partial m} &= - 
  \frac{(l e^{\phi})}{16 \pi^2 r^4 \kappa \epsilon \sqrt{1 - 2m/r}} 
  \label{eq:coeq2} 
  \, ,
\end{align}
where the radial distance $r$, mass $m(r)$, and metric function $\phi(r)$ are macroscopic parameters (obtained by solving the TOV equations). The
specific heat $c_v(r,T)$, thermal conductivity $\kappa(r,T)$, neutrino emissivity $Q(r,T)$, and energy density $\epsilon(r)$ are microscopic parameters. Solving~\Cref{eq:coeq1} and~\Cref{eq:coeq2} determines the temperature $T(r,t)$ and luminosity $l(r,t)$. 

The specific heat is typically dominated by neutrons and protons in the core and by neutrons and ions in the crust. Pairing, if present, must also be taken into account, as it strongly suppresses the specific heat of nucleons below the critical temperature. A detailed description of the specific heat of neutron star matter can be found in Refs.~\cite{Yakovlev2000,Yakovlev2004}. 
The thermal conductivity is dominated by the scattering of degenerate electrons, both in the core and in the crust. A review of the thermal conductivity calculations relevant for neutron star cooling can be found in Refs.~\cite{gnedin1995thermal,potekhin2015neutron}.

\subsubsection{Neutrino Emissivity}
\label{sec:nu_emissivity}

In order to solve~\Cref{eq:coeq1} and~\Cref{eq:coeq2}, we need to compute the neutrino emissivity. We consider three types of processes that contribute to the total neutrino emissivity in the neutron star core: direct Urca, modified Urca, and nucleon bremsstrahlung~\cite{Friman1979NeutrinoEmission,Lattimer1991DirectUrca,Itoh1996NeutrinoProcesses,Yakovlev2001NeutronStarCooling}. We emphasize that the direct and modified Urca processes are considered for the entire baryon octet and its parity partners. For the crust, we consider pair creation/annihilation, nucleon bremsstrahlung, and plasmon decay~\cite{Friman1979NeutrinoEmission,Itoh1996NeutrinoProcesses,Yakovlev2001NeutronStarCooling}. Since our model includes quarks, we also include the corresponding quark neutrino emission processes~\cite{Iwamoto1982QuarkCooling}.

Following the work of Ref.~\cite{Brodie:2025nww}, the novel feature of this study is the inclusion of the direct Urca process for the parity partners
\begin{align}
    &\neutronMinus \rightarrow \proton + e + \bar{\nu}_e &&\proton + e \rightarrow \neutronMinus + \nu_{e} \label{eq: n^*d}\\
    &\neutron \rightarrow \protonMinus + e + \bar{\nu}_e &&\protonMinus + e \rightarrow \neutron + \nu_{e} \label{eq: nd(p^*)}\\
    &\neutronMinus \rightarrow \protonMinus + e + \bar{\nu}_e  &&\protonMinus + e \rightarrow \neutronMinus + \nu_{e}\,. \label{eq: n^*d(p^*)}
\end{align}
We only show processes involving the nucleon parity partners (and focus the rest of this section on them) because the parity partners of the hyperons do not appear below the central density of the most massive star predicted by the PD-CMF model.

We compute the direct Urca neutrino emissivity using Equation~(120) of Ref.~\cite{Yakovlev:2000jp} and Equation~(7) of Ref.~\cite{Lattimer:1991ib}. 
At low temperatures, $T \lesssim 1$~MeV, the emissivity is dominated by particles near their Fermi surfaces. The sum of emissivities for each pair of neutron decay and electron capture direct Urca (dU) processes is
\beq
Q^{\text{dU}}=\frac{457\ \pi}{10080} G_F^2 \cos^2{\theta_C}(1+3\,g_{A}^2)m^*_{\neutrons} m^*_{\protons} m_e T^6 \Theta^{\text{dU}}.
\label{eq: du_emissivity}
\eeq
Here $G_F = 1.16637 \times 10^{-11}$ MeV$^{-2}$, 
the Cabibbo angle $\theta_C = 13.02^{\circ}$, and $g_{A}= 1.267$.
Because of the uncertainty in the axial vector couplings of the nucleon parity partners, we set them equal to $g_A=1.267$. The in-medium particle masses are denoted $m_{\pm}$. $\Theta^{\text{dU}}=1$ if the direct Urca process is kinematically 
allowed and $\Theta^{\text{dU}}=0$ if not. 

At low temperatures, the direct Urca process is allowed if the following kinematic relations are obeyed for the Fermi momenta
\begin{align}
    k^{F}_{\protons}+&k^{F}_{e^-}\geq k^{F}_{\neutrons}\,,\\ k^{F}_{\neutrons}+&k^{F}_{\protons}\geq k^{F}_{e^-}\,,\\ k^{F}_{e^-}+&k^{F}_{\neutrons}\geq k^{F}_{\protons}\,.
\end{align}
When direct Urca processes
are not allowed, modified Urca processes become the dominant neutrino emission channel. These are given by
\begin{align}
\label{eq: mu_reactions}
    \neutron + \nucleon &\rightarrow \proton + \nucleon + e^{-} + \bar{\nu}_e\,, \nonumber \\
    \proton + \nucleon + e^{-} &\rightarrow \neutron + \nucleon + \nu_e \; ,
\end{align}
where we use $N$ to mean either a neutron or a proton. When $\nucleon=\neutron$ in~\Cref{eq: mu_reactions}, the modified Urca contribution to the emissivity 
is given by Equation~(140) in Ref.~\cite{Yakovlev:2000jp} and Equation~(65c) in Ref.~\cite{Friman:1979ecl}
\beq
Q^{\text{mU,}\neutron} = A\, G_F^2 \cos^2{\theta_C} g_A^2 
\frac{(m^{*}_{\neutron})^3 m^{*}_{\proton} k^{F}_{\proton}}{m_{\pi}^4} T^8 \; ,
\eeq
where $m_{\pi} = 139$~MeV is
the pion mass and the constant $A = 0.04656$ \cite{Friman:1979ecl}.
When $\nucleon = \proton$ in~\Cref{eq: mu_reactions}, the modified Urca contribution to the emissivity follows from Equation~(142) of Ref.~\cite{Yakovlev:2000jp}
and is
\beq
Q^{\text{mU,}\proton} \approx Q^{\text{mU,}\neutron} \frac{(m^*_{\proton})^2}{(m^*_{\neutron})^2} \frac{(k^{F}_{e^-}+3k^{F}_{\proton}-k^{F}_{\neutron})^2}{8 k^{F}_{e^-}\ k^{F}_{\proton}} \, \Theta^{\text{mU,}\proton},
\eeq
where $\Theta^{\text{mU,}\proton}$ is 1 if $k^{F}_{e^-}+3k^{F}_{\proton}>k^{F}_{\neutron}$ and 0 otherwise. 

In addition to the Urca processes for the baryons, Urca processes for quark matter are also considered. Their neutrino emissivities follow closely that of the nucleons and are described in Refs.~\cite{Lattimer1991DirectUrca,Yakovlev2001NeutronStarCooling}.
\subsubsection{Pairing}
\label{sec: pairing}
We also incorporate nucleon pairing into our thermal evolution calculations. Pairing plays a crucial role in accurately simulating the cooling of neutron stars \cite{ho2015tests}, as it suppresses the neutrino emissivity of baryonic processes by
\begin{equation}
    Q = Q_0R\,,
\end{equation}
where $Q_0$ is the unpaired emissivity and $R\leq1$ is the suppression factor. The actual value of the suppression factor will depend on the pairing gap, which in turn will depend on the temperature, Fermi momentum, and pairing model, as described in Ref.~\cite{Yakovlev2000}.

\begin{figure}[t]
\centering
\includegraphics[width=0.5\textwidth]{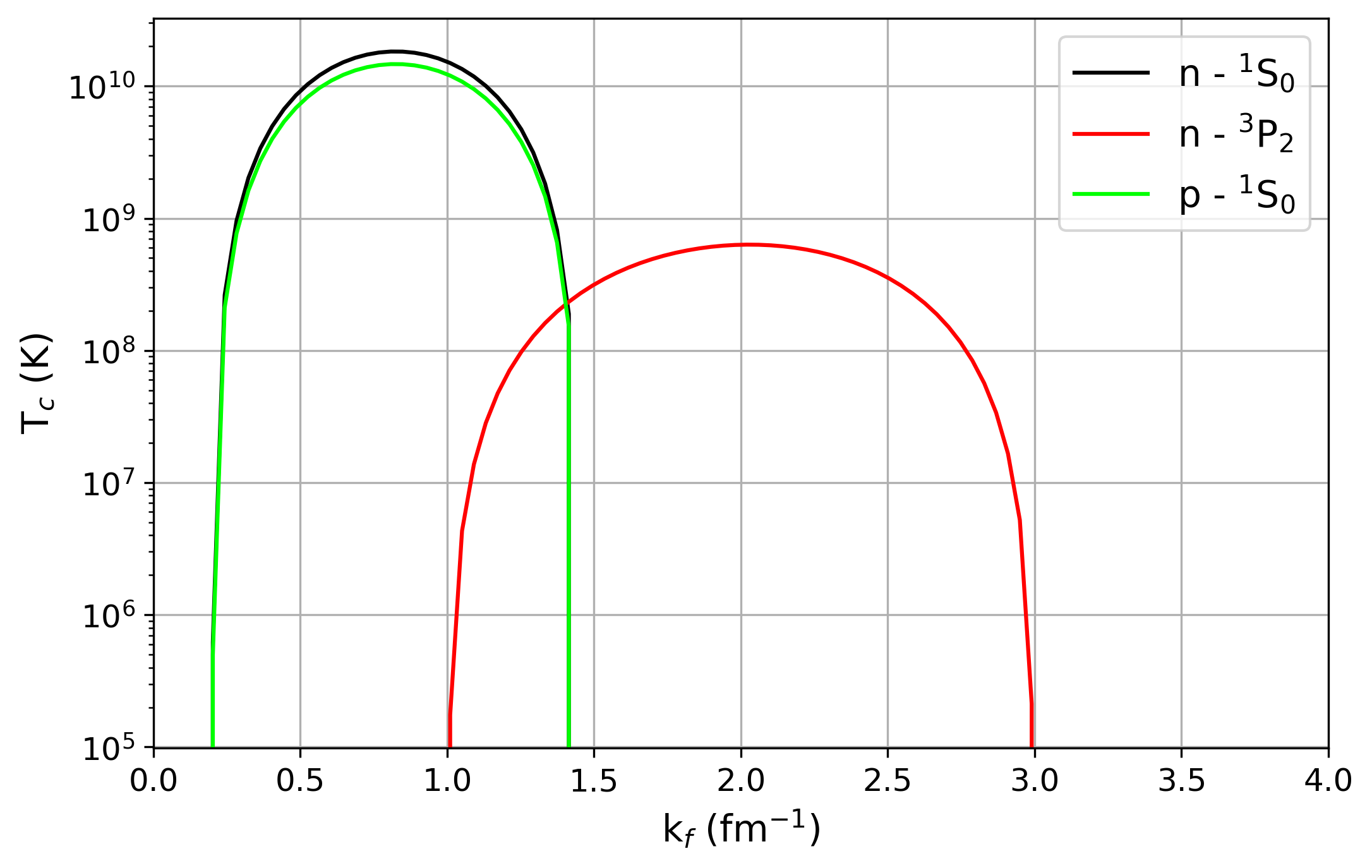}
\caption{Critical temperature as a function of Fermi momentum for the different pairing patterns assumed to take place in the neutron star. }
\label{fig:pairTC}
\end{figure}

For this work, we use the CCDK model \cite{chen1993pairing} to describe both neutron and proton singlet pairing ($^1S_0$) and the AO model \cite{amundsen1985superfluidity} for neutron triplet pairing ($^3P_2$).  Both models have been applied as outlined in \cite{ho2015tests}. We show in~\Cref{fig:pairTC} the critical temperature as a function of the Fermi momentum for the singlet and triplet pairing patterns.

In principle, pairing of the parity partners to the nucleons could also be considered
(cross pairing),
$\langle \nucleon \nucleonMinus \rangle$. This pairing pattern is suppressed because the Fermi momentum for $\nucleon$ and
$\nucleonMinus$ differ for the relevant densities considered.
Pairing of the nucleon parity partners $\nucleonMinus$ to themselves,
$\langle \nucleonMinus \nucleonMinus \rangle$ could also be considered. 
Such pairing
presumably occurs, but as the Fermi momentum of the $\nucleonMinus$
is much smaller than the $\nucleon$ for the density range where the $\nucleonMinus$ neutrino emissivity is nonzero, presumably so are the $\nucleonMinus$ gaps (see Equation~(11) in Ref.~\cite{Sedrakian:2018ydt} and Equation~(18) in Ref.~\cite{Page:2013hxa}). Assuming that the $\nucleonMinus$ gap is small compared to the $\nucleon$ gap
implies that neutrino emission due to $\neutronMinusDecayProton$ is even larger compared to standard neutron decay $\neutronDecayProton$, which is
suppressed by $\nucleon$ pairing.

In this work, we do not include the possibility of hyperon pairing. Although a few studies have examined hyperonic pairing, particularly for $\Lambda$ and $\Xi$ hyperons (see, for example, Refs.~\cite{Anzuini:2021rjv, raduta2018cooling,PhysRevD.103.083004,Raduta:2019rsk,Grigorian:2018bvg,Fortin:2021umb}), large uncertainties remain. Moreover, due to the small hyperon population predicted by our model and their negligible impact on the thermal evolution, hyperon pairing is expected to have little to no influence on our results. See also Refs.~\cite{Raduta:2017wpp,Grigorian:2018bvg,Raduta:2019rsk,Fortin:2021umb} for previous neutron star cooling studies involving hyperons.

The PD-CMF model also incorporates deconfined quarks. Quarks affect the neutrino emissivity through direct and modified quark Urca processes~\cite{IWAMOTO19821,PhysRevLett.44.1637}. We explore some aspects of quark pairing \cite{PhysRevD.66.063003,Page:2004fy}, but
due to the uncertainties regarding quark matter at high densities, there is no consensus on the preferred pairing pattern for quark matter. Most studies, however, consider the color-flavor-locked (CFL) phase \cite{Alford:1998mk,Alford:2007xm} or two-flavor color-superconductivity (2SC)~\cite{Alford:1998mk,Rapp:1997zu,grigorian2005cooling}. For our study, we consider the CFL pairing pattern, where all quarks of every color are paired and therefore have their neutrino emissivities exponentially suppressed. We assume that the critical temperature for the quarks is $T_c = 0.4\,\Delta$, with a gap of $\Delta = 10$ MeV~\cite{Blaschke2001}. 

\subsection{Atmosphere and Envelope Modeling}
The thermal radiation observed from cooling neutron stars emerges from a thin surface atmosphere, while the insulating outer envelope determines the relation between the interior and surface temperature. The envelope typically has a thickness of $30–100\,\rm{m}$~\cite{Gudmundsson1982} and is therefore treated as a boundary layer in cooling simulations. The specific boundary conditions depend on the envelope composition. Because light elements have higher thermal conductivity, they generally make the envelope less insulating. In this work, we consider two cases for the composition of the envelope: an envelope containing only heavy elements, corresponding to an envelope mass of $\Delta M/M = 10^{-18}$, and another composed of a carbon layer $\Delta M/M = 10^{-8}$, where $\Delta M$ denotes the envelope mass and $M$ is the total mass of the star. The relationship between the temperature of the interior and surface is given in Ref.~\cite{Potekhin_2003}.

\section{Results}

\begin{figure}[t!]
\centering
\includegraphics[width=0.5\textwidth]{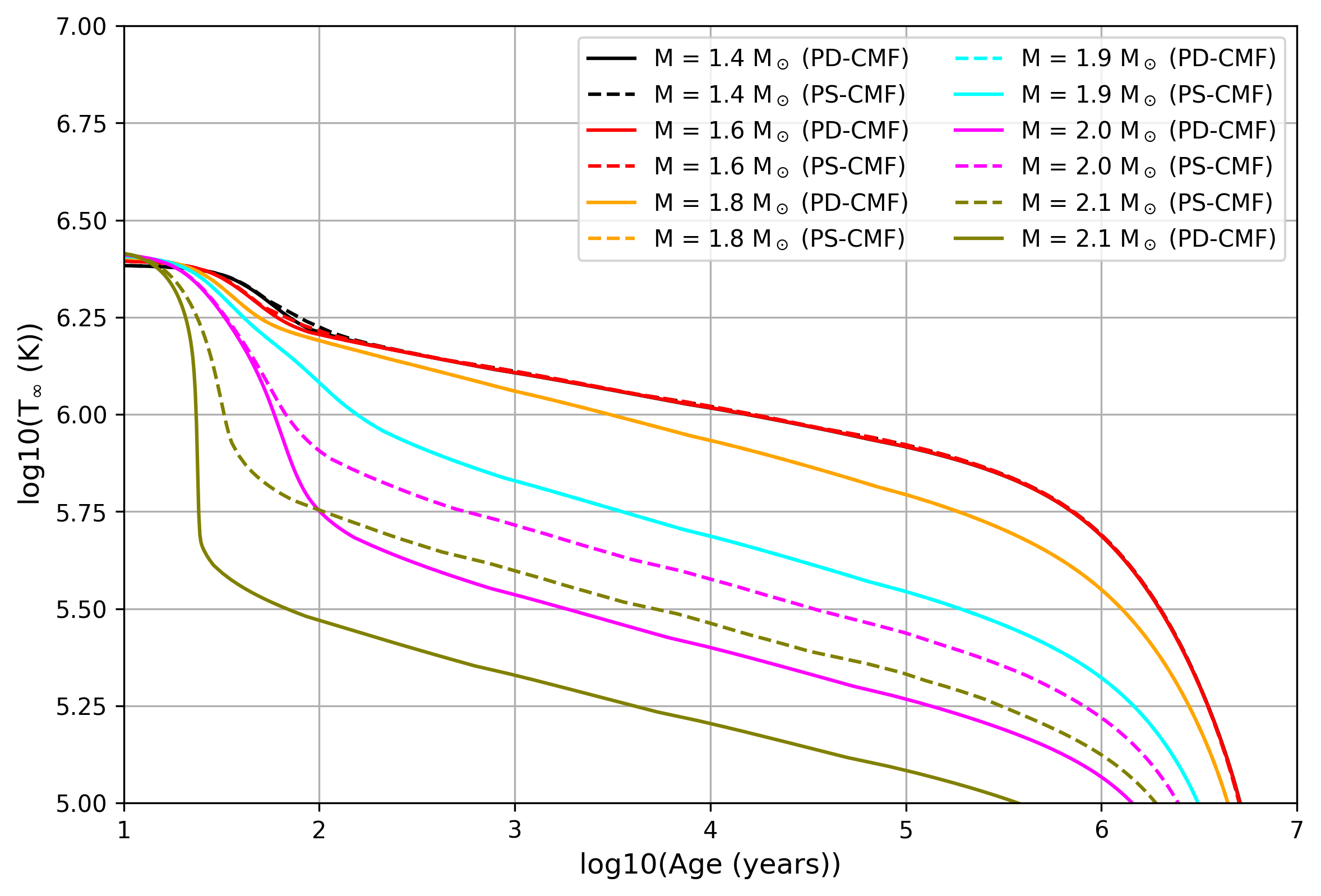}
\caption{Neutron star thermal evolution for stars of different masses from the model with the nucleon parity partners (PD-CMF, dashed lines) and without (PS-CMF, solid lines). We show the redshifted surface temperature measured by an observer at infinity as a function of time after formation.}
\label{fig:cool_1}
\end{figure}

\subsection{Unpaired Matter}
We begin our analysis by comparing the thermal evolution of neutron stars with and without the nucleon parity partners to quantify their impact on cooling, see~\Cref{fig:cool_1}. 
We consider a standard, heavy-element envelope and do not consider pairing in this subsection. More realistic cooling scenarios will be discussed in the following subsections. 

\Cref{fig:cool_1} demonstrates that low-mass stars, where the nucleon parity partners do not appear in the PD-CMF model, display nearly identical thermal evolution to those of the PS-CMF model, as expected. Higher mass stars, where the nucleon parity partners appear in the PD-CMF model, cool more rapidly. As predicted by Ref.~\cite{Brodie:2025nww}, this rapid cooling is attributed to the onset of the direct Urca processes involving the nucleon parity partners. 

\begin{figure}[t!]
\centering
\includegraphics[width=0.5\textwidth]{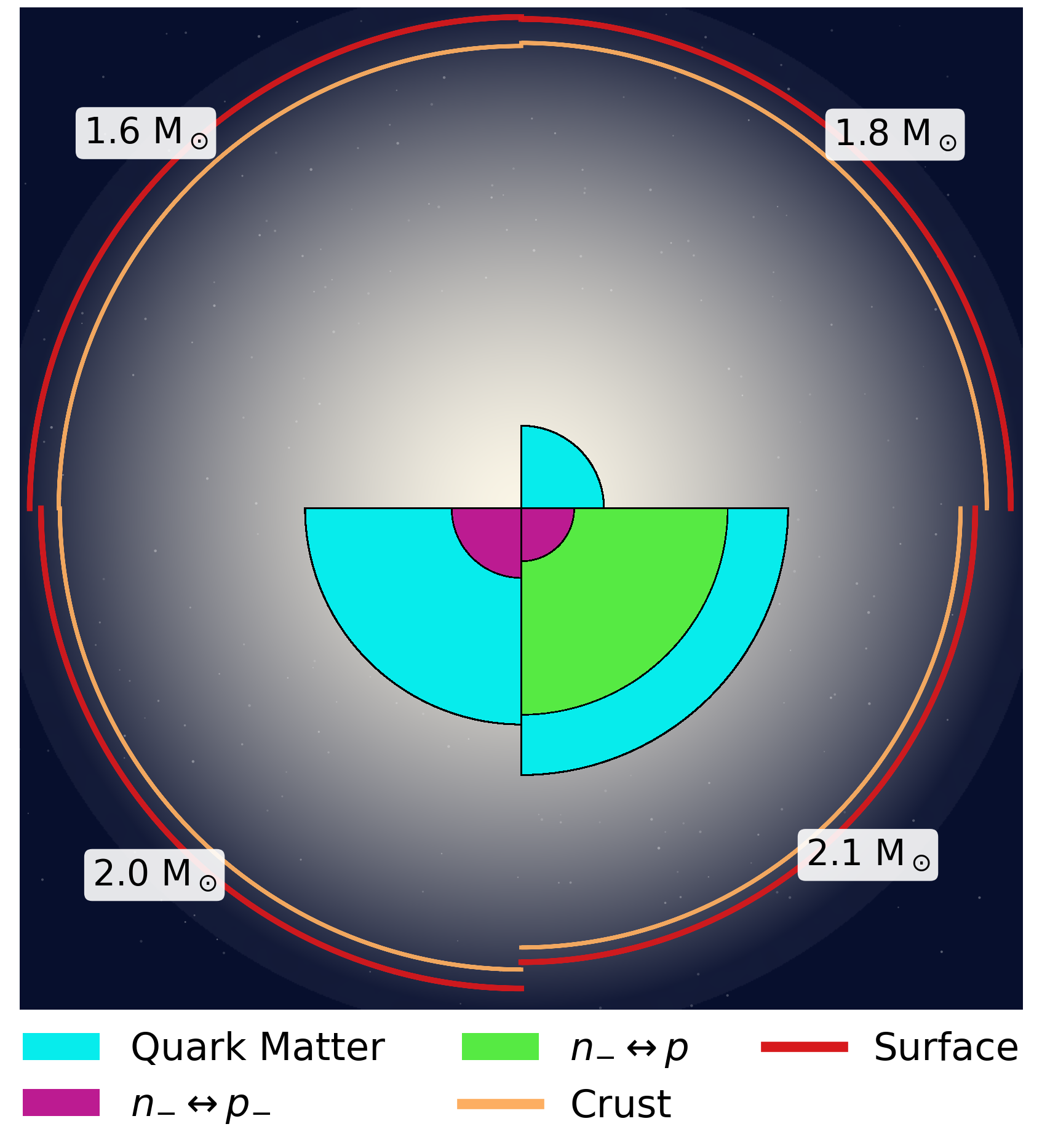}
\caption{Cross sections of neutron stars indicating the regions where the direct Urca processes involving the nucleon parity partners~(\Cref{eq: n^*d} and~\Cref{eq: n^*d(p^*)}) are kinematically allowed inside stars of different masses as described by the PD-CMF model. For comparison, the location in the stars where quark matter is present is shown as well.  
}
\label{fig:cross_section}
\end{figure}

In~\Cref{fig:cross_section} we show the regions where the nucleon parity partner direct Urca processes (\Cref{eq: n^*d} and \Cref{eq: n^*d(p^*)}) are allowed for stars of different masses. The lowest mass stars shown do not allow any parity partner direct Urca processes, while for the $2\,\rm{M}_\odot$ star, the $\neutronMinus\leftrightarrow \protonMinus$ channel (\Cref{eq: n^*d(p^*)}) is allowed. The $2.1\,\rm{M}_\odot$ star allows both the $\neutronMinus\leftrightarrow \protonMinus$ (\Cref{eq: n^*d(p^*)}) and $\neutronMinus\leftrightarrow \proton$ (\Cref{eq: n^*d}) processes. For reference,~\Cref{fig:cross_section} also shows the region in which quark matter is found. As shown in~\Cref{fig:cool_1}, the $2\,\rm{M}_\odot$ PD-CMF star cools much faster than the PS-CMF star, even though there is only a small fraction of the star where the $\neutronMinus\leftrightarrow \protonMinus$ process is allowed, see the lower left quadrant of~\Cref{fig:cross_section}. 

We have included the contribution of the nucleon parity partners to the specific heat and found it to be insignificant compared to the nucleons. The primary way the nucleon parity partners affect the thermal evolution is through their direct Urca processes.

\subsection{Paired Nuclear Matter}
\begin{figure*}[tbp]
\centering
\includegraphics[width=0.329\textwidth]{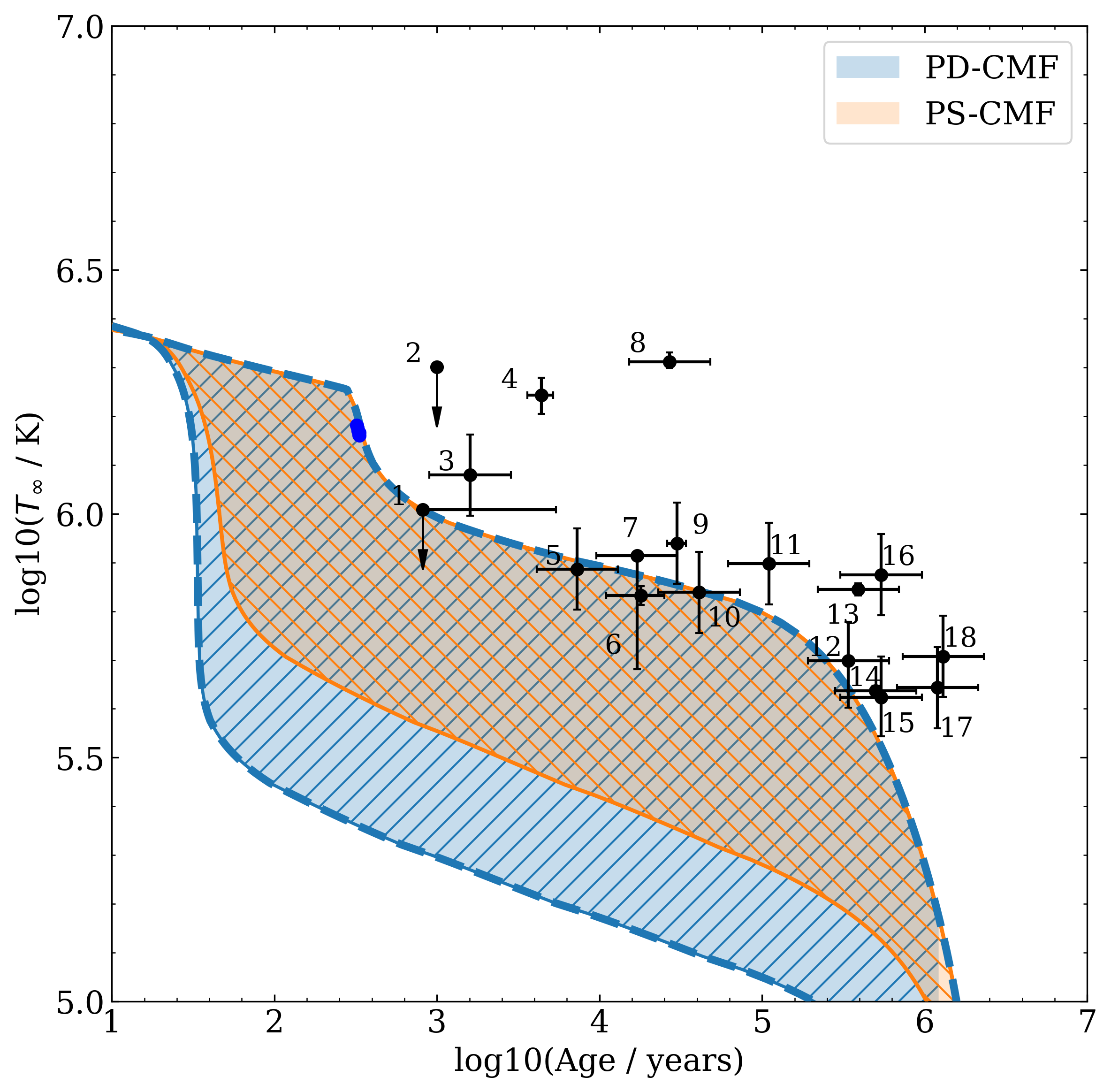}
\includegraphics[width=0.329\textwidth]{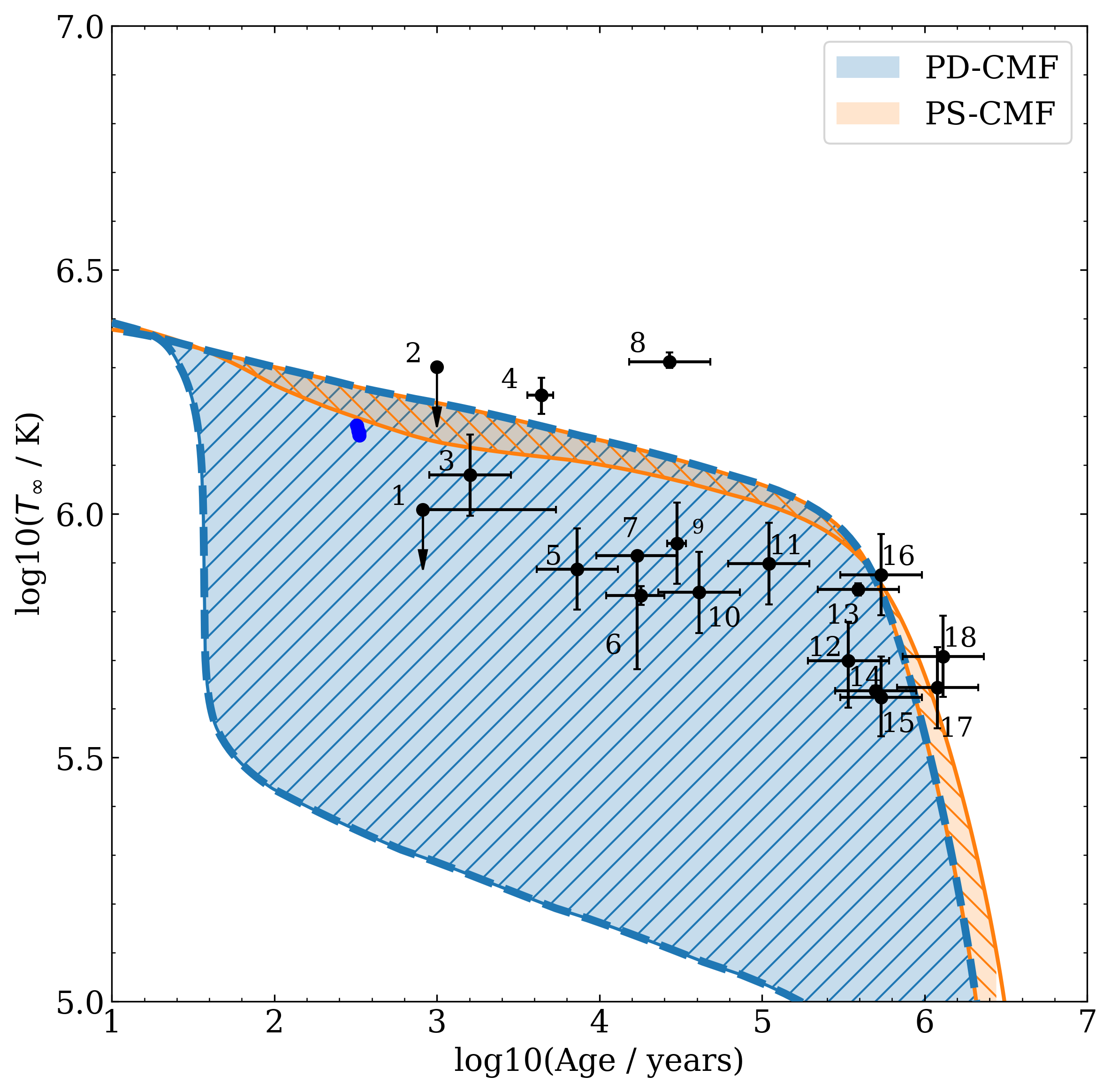}
\includegraphics[width=0.329\textwidth]{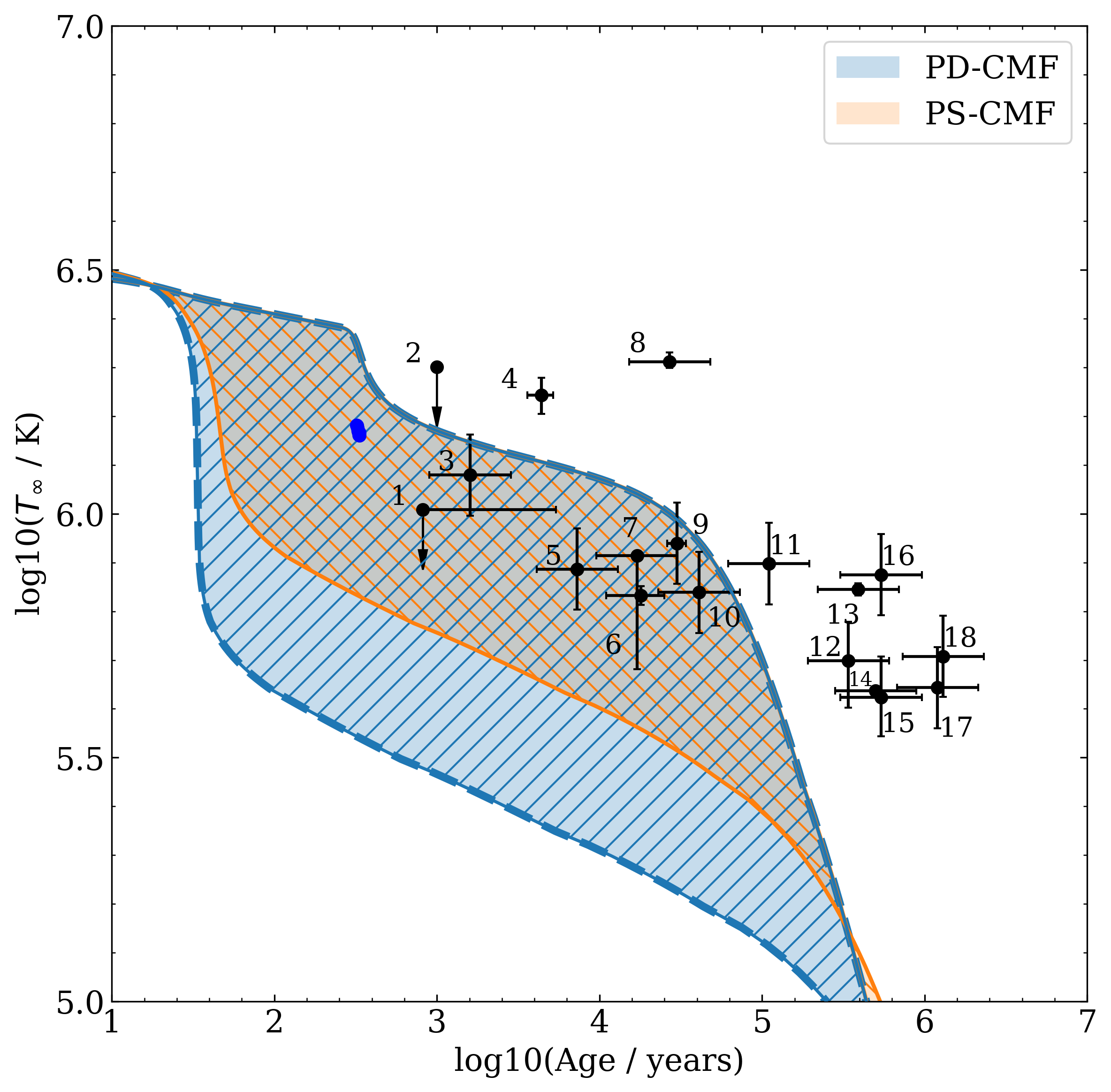}
\caption{The thermal evolution of neutron stars under various assumptions compared with astrophysical observations of isolated, middle-aged neutron stars.\\
(a) Left: Cooling bands from the simulation representing the thermal evolution for stars with masses between $1.4\,\rm{M}_\odot-2.1\,\rm{M}_\odot$. The blue (orange) shaded region is from the model with (without) the nucleon parity partners. The nucleons are the only paired particle species.
Observed data is from Ref.~\cite{Beznogov:2014yia}, also see~\Cref{table}.\\
(b) Middle: Same as the left panel but quarks as well as nucleons are paired.\\ (c) Right: Same as the left panel except for stars with an envelope of light, instead of heavy, elements with an envelope mass $\Delta M/M = 10^{-8}$.}
\label{fig:cool}
\end{figure*}

In the left panel of~\Cref{fig:cool} we present results obtained by incorporating pairing for the nucleons (not their partners, see Sec.~\ref{sec: pairing}), and by comparing the thermal evolution of stars with and without parity partners. The shaded bands indicate the region between the cooling curves of a $1.4\,\rm{M}_\odot$ star (upper limit) and a $2.1\,\rm{M}_\odot$ star (lower limit), with intermediate masses in between. The cooling band excluding the nucleon parity partners is shown in orange and the band including them is shown in blue. 

Even though the nucleons are paired, the cooling pattern in the left panel of~\Cref{fig:cool} mirrors the qualitative behavior observed in~\Cref{fig:cool_1}, where more massive stars that contain the nucleon parity partners cool more rapidly. The addition of nucleon pairing should increase the difference between the PD-CMF and PS-CMF cases because neutrino emission from nucleon-based processes is suppressed, while neutrino emission from the nucleon parity partners is unaffected. However, the dominant contribution to the neutrino emissivity in the PS-CMF model comes from quark direct Urca, e.g., processes like $d\rightarrow u + e + \bar{\nu}_e$ and $u + e \rightarrow d + \nu_e$. This is because nucleon direct Urca is not kinematically allowed and is why there is no qualitative change to the PS-CMF band compared to~\Cref{fig:cool_1} even though the nucleons are paired.

To compare our cooling simulations with observed neutron star temperatures, we use the cataloged sources from Ref.~\cite{Beznogov:2014yia}, which are listed in~\Cref{table}. For entries with available uncertainties, we adopt 2$\sigma$ error bars. If there are no error bars available, we assume a variation factor of 0.5 to 2 in both temperature and age. Note that source 8 – XMMU J1731-347 from Ref.~\cite{Beznogov:2014yia} has been replaced with HESS J1731-347, as reported in Ref.~\cite{Doroshenko:2022}, reflecting updated measurements that indicate a modest increase in surface temperature. We find a moderate disagreement between our cooling simulations and the observational data. However, we have yet to consider the pairing of quark matter; we do so in the next subsection.

\begin{table}[t!]
\centering
\begin{tabular}{cl}
\hline
\textbf{ID} & \textbf{Object} \\
\hline
0  & CasA NS \\
1  & PSR J0205+6449 (3C58) \\
2  & PSR B0531+21 (Crab) \\
3  & PSR J1119$-$6127 \\
4  & RX J0822$-$4300 (PupA) \\
5  & PSR J1357$-$6429 \\
6  & PSR B1706$-$44 \\
7  & PSR B0833$-$45 (Vela) \\
9  & PSR J0538+2817 \\
10 & PSR B2334+61 \\
11 & PSR B0656+14 \\
12 & PSR B0633+1748 (Geminga) \\
13 & PSR J1741$-$2054 \\
14 & RX J1856.4$-$3754 \\
15 & PSR J0357+3205 (Morla) \\
16 & PSR B1055$-$52 \\
17 & PSR J2043+2740 \\
18 & RX J0720.4$-$3125 \\
\hline
\end{tabular}
\caption{A catalog of 18 isolated, middle-aged pulsars from Ref.~\cite{Beznogov:2014yia} whose observed or inferred ages and temperatures are compared with our cooling simulations.}
\label{table}
\end{table}

\subsection{Paired Quark Matter}

The middle panel of \Cref{fig:cool} shows the cooling bands including quark pairing (as well as nucleon pairing), see Sec.~\ref{sec: pairing}. In stars without nucleon parity partners, the cooling band significantly narrows, i.e., stars with masses between $1.4\,\rm{M}_\odot-2.1\,\rm{M}_\odot$ cool slowly, leading to disagreement with the observed data. The narrowing arises from the substantial suppression of neutrino emission from processes involving quarks. For stars described by the PD-CMF model, low-mass stars experience notably slow cooling due to suppressed neutrino emission from quarks, whereas more massive stars still cool rapidly, owing to the unsuppressed neutrino emission from the nucleon parity partners. This results in the cooling band being broad enough to accommodate most observations. 

\subsection{Impact of Atmospheric Modeling}

In the left panel of~\Cref{fig:cool}, some observed thermally emitting neutron stars are consistent with the upper part of the cooling bands we show, but most observations are stars with temperatures higher than those predicted by our cooling simulations. These cooling simulations were done under the assumption of a heavy-element neutron star envelope, which is associated with a lower surface temperature compared to an envelope of light elements, because it is more insulating. We show the impact of a light-element-rich envelope in the right panel of~\Cref{fig:cool}. We use an envelope with $\Delta M/M = 10^{-8}$, where $\Delta M$ represents the mass of light elements in the outer envelope \citep{Gudmundsson1982,Gudmundsson1983,Potekhin1997}. Although this atmosphere model improves the agreement of the simulation cooling bands with the observation, compared to the left panel, the cooling bands shown in the right panel still cannot describe a number of older sources, which remain significantly hotter than predicted by the models.

\section{Conclusions}
\label{sec:conclusions}

Examining the thermal evolution of neutron stars allows us to investigate the microphysics of dense matter. Dense matter lies in a strong-coupling regime where QCD cannot yet be solved. To describe it, we make use of the phenomenological parity doublet chiral
mean field (PD-CMF) model, which describes the interactions between the entire baryon octet, i.e., nucleons and hyperons, as well as the parity partners of all the baryons.

We performed cooling simulations of neutron stars for a range of masses. We found that neutrino emission from direct Urca processes involving the nucleon parity partners significantly impacts the thermal evolution of massive stars. The stars that contain a finite population of nucleon parity partners cool more rapidly than their counterparts without the parity partners. In the PD-CMF model we used, the parity partners of the hyperons do not appear below the central density of the heaviest mass star predicted by the model. 

Although the spatial region in which the direct Urca processes involving the nucleon parity partners are allowed may be limited to the innermost core, their neutrino emissivity is the dominant contribution. As a result, they can drive rapid cooling in sufficiently massive stars. The quantitative impact depends on the density at which parity partners appear, which is controlled by the model Lagrangian and coupling constants.

We investigated how variation in the composition of the neutron star envelope and pairing in both nuclear and quark matter affects the thermal evolution as well. In the case of a heavy-element envelope (the standard assumption) and including pairing of the nucleons and of the quarks, we found that the model with parity partners is more consistent with most of the astrophysical observations of neutron star surface temperatures and ages. Some stars are not described by any of the variations we consider, e.g., the hotter-than-average neutron star, XMMU J1731-347, whose unusually high temperature has been examined in detail, for example, in Refs.~\cite{Zhang_2025,sagun2023nature}. 

We did not take into account the possibility of pairing between the nucleon parity partners, which, if the pairing gap is large enough, could qualitatively impact the neutron star thermal evolution. If sizable pairing gaps were present for the parity partners, the associated direct Urca processes would be exponentially suppressed, potentially eliminating the rapid-cooling channel identified in this work. Due to their small Fermi momenta compared to the positive parity nucleons, we do not expect their pairing gap to be as large. To date, the pairing of the nucleon parity partners has not been addressed in the literature.

Future work incorporating pairing of the parity partners and exploring alternative chiral parameterizations will help determine whether rapid cooling induced by parity doubling is a robust and observable signature of chiral symmetry restoration in dense matter.

\section{Acknowledgments}

LB is supported by the U.S. Department of Energy, Office of Science, Office of Nuclear Physics, under Award No.~\#DE-FG02-05ER41375.
V.D. acknowledges support from the Department of Energy under grant DE-SC0024700 and from the National Science Foundation under grants MUSES OAC2103680 and NP3M PHY2116686. 
R.D.P. is supported by the U.S. Department of Energy under contract DE-SC0012704, and by the Alexander von Humboldt Foundation. 

\appendix 
\counterwithin{figure}{section}
\counterwithin{table}{section}
\section{Parity Doublet Model}
\label{App:CMF}

In the parity doublet chiral mean field (PD-CMF) model, the Lagrangian describes scalar and vector interactions among the ground-state baryons with spin $1/2$ and their parity partners. We solve for the fields under the mean-field approximation by extremizing the thermodynamic potential density (or, equivalently, maximizing the total pressure) with respect to each field while ensuring charge neutrality and chemical equilibrium. The Lagrangian ${\cal L}_{{\rm SU}(3)}$ consists of the following parts (presented also in Ref.~\cite{Motornenko:2020yme})
\begin{align}
{\cal L}_{\rm{SU(3)}}={\cal L_B}  + U_{\rm scalar} + U_{\rm vector} \,.
\end{align}
The interaction term is
\begin{align}
{\cal L_B} &= \sum_b (\bar{B_b} i \slashed{\partial} B_b)
- \sum_b  \left(\bar{B_b} m^*_b B_b \right) \nonumber \\ &+
\sum_b  \left(\bar{B_b} \gamma_\mu (g_{\omega b} \omega^\mu +
g_{\rho b} \rho^\mu + g_{\phi b} \phi^\mu) B_b \right) \,,
\label{lagrangian2}
\end{align}
where the index $b$ runs through the octet baryons and their respective parity partners.

The dynamics of chiral fields, associated with chiral symmetry (especially its spontaneous breaking), are determined self-consistently by the potential of the scalar fields $\sigma$ and $\zeta$, 
\begin{align}
U_{\rm scalar} & =  V_0 - \frac{1}{2} k_0 I_2 + k_1 I_2^2 - k_2 I_4 + k_6 I_6   \nonumber \\
& + k_4 \ln{\frac{\sigma^2\zeta}{\sigma_0^2\zeta_0}} - U_{\rm sb} \,,
\label{veff}
\end{align}
with
\begin{align}
    I_2 &= (\sigma^2+\zeta^2)\,,\nonumber\\
    I_4 &= -(\sigma^4/2+\zeta^4)\,,\nonumber\\
    I_6 &= (\sigma^6 + 4\, \zeta^6)\,,
\end{align} 
where $V_0$ ensures that the pressure in vacuum vanishes. The terms $I_n$ correspond to the possible chiral invariants that form different meson-meson interactions. The logarithmic term in~\Cref{veff}, introduced in Refs.~\cite{Heide:1993yz,Papazoglou:1996hf}, contributes to the QCD trace anomaly and is motivated by the form of the QCD beta function at the one-loop level. In addition, an explicit symmetry-breaking term is introduced in the scalar potential:
\begin{equation}
U_{\rm sb} = m_\pi^2 f_\pi\sigma +\left(\sqrt{2}m_K^ 2f_K-\frac{1}{\sqrt{2}}m_\pi^ 2 f_\pi\right)\zeta\,,
\label{vsb}
\end{equation}
where $f_\pi$ and $f_K$ are the pion and kaon decay constants, respectively.

\begin{table}[t!]
\centering
\begin{tabular}{|c|c|>{\columncolor[gray]{0.1}}c|c|c|>{\columncolor[gray]{0.1}}c|c|c|}
 \hline
        $m_{\pi}$ & 138 MeV & \ \ & $g_{q\sigma}$ & -1 & \ \ & $g_{\rho p,n}$ & $\pm4.9$ \\ \hline
        $m_{K}$ & 498 MeV &  & $g_{s\zeta}$ & -1 &  & $g_{\rho \Lambda}$ & 0 \\ \hline
        $m_{\omega}$ & 783 MeV &  & $g_{\sigma N}^{(1)}$ & -9.81 &  & $g_{\rho{\Sigma^{\pm}}}$ & $\pm3.63$ \\ \hline
        $m_{\rho}$ & 761 MeV &  & $g_{\sigma \Lambda}^{(1)}$ & -7.98 &  & $g_{\rho {\Xi^{\pm}}}$ & $\pm1.816$ \\ \hline
        $m_{\phi}$ & 1019 MeV &  & $g_{\sigma \Sigma}^{(1)}$ & -5.83 &  & $g_{\phi N}$ & 0 \\ \hline
        $m_{0q}$ & 225 MeV &  & $g_{\sigma \Xi}^{(1)}$ & -4.89 &  & $g_{\phi \Lambda}$ & -3.34 \\ \hline
        $\delta m_q$ & 350 MeV &  & $g_{\sigma B}^{(2)}$ & 3.21 &  & $g_{\phi \Sigma}$ & -3.34 \\ \hline
        $ m_s$ & 130 MeV &  & $g_{\zeta N}^{(1)}$ & -1.15 &  & $g_{\phi \Xi}$ & -6.69 \\ \hline       
        $f_{\pi}$ & 93 MeV &  & $g_{\zeta \Lambda}^{(1)}$ & -3.74 &  & $k_0$ & 242$^2$ $\mathrm{MeV}^2$ \\ \hline
        $f_{K}$ & 122 MeV &  & $g_{\zeta \Sigma}^{(1)}$ & -6.02 &  & $k_1$ & 4.818 \\ \hline
        $m_0$ & 675 MeV &  & $g_{\zeta \Xi}^{(1)}$ & -7.35 &  & $k_2$ & 23.3 \\ \hline
        $\zeta_0$ & -106.77 MeV &  & $g_{\zeta B}^{(2)}$ & 0 &  & $k_4$ & 76$^4$ $\mathrm{MeV}^4$ \\ \hline
        $\sigma_0$ & -93.0 MeV &  & $g_{\omega N}$ & 7.02 &  & $k_6$ & $10^{-4}$ $\mathrm{MeV}^{-2}$ \\ \hline
        $V_0$ & $-(229^4)\,\mathrm{MeV}^{4}$ &  & $g_{\omega \Lambda}$ & 4.84 &  & $\beta_2$ & 1500 \\ \hline
         &  &  & $g_{\omega \Sigma}$ & 8.175 &  & $Z_{\phi}$ & 2.239 \\ \hline
         &  &  & $g_{\omega \Xi}$ & 4.905 &  & $Z_{\omega}$ & 1.322 \\ \hline

\end{tabular}
\caption{List of parameters and coupling constants of the PD-CMF model. $q$ indexes $u,d,s$ quark flavors, $N$ over positive and negative parity nucleons, and $B$ over positive and negative parity baryons. Here $\Sigma^{\pm}$ and $\Xi^{\pm}$ refer to their electric charges, not their parity states.}
\label{param-table}
\end{table}

The mean-field vector interaction at finite baryon density, finite isospin density, and finite strangeness density depends on the respective conserved charge densities and is controlled by the vector $\omega$, $\rho$, and $\phi$ fields in the vector potential
\begin{align}
U_{\rm vector}&= -\frac12\left(m_\omega^2\omega^2 + m_\rho^2\rho^2 + m_\phi^2\phi^2\right)\nonumber\\ &-g_4\left(\omega^4+6\beta_2\omega^2 \rho^2+ \rho^4+                 \frac12\phi^4\left(\frac{Z_\phi}{Z_\omega}\right)^2 \right.\nonumber\\
&+3\left.\left(
\rho^2+\omega^2\right)\left(\frac{Z_\phi}{Z_\omega}\right)\phi^2\right)\,.
\end{align}
This potential is based on the C3 chiral invariant for self-interactions \cite{Dexheimer:2015qha} with an adjustable parameter $\beta_2$ for the vector-isovector term, important for reproducing astrophysical constraints \cite{Dexheimer:2018dhb}. $Z_\omega$ and $Z_\phi$ are constants related to a vector field redefinition that lifts the degeneracy between the $\omega$ and $\phi$ fields (see Ref.~\cite{Kumar:2024owe} for details).

All fixed parameters and coupling constants used in the PD-CMF model are summarized in~\Cref{param-table}.
The vacuum masses of all baryons and their parity partners in the model are shown in~\Cref{tab:masses}. We do not have any isospin symmetry-breaking interactions for the masses included in the present version of the model.

\begin{table}[t!]
\centering
\begin{tabular}{|c|c|c|}
 \hline
        Baryon & Ground state mass $[\mathrm{MeV}]$ & Parity partner mass $[\mathrm{MeV}]$ \\ \hline
        N & 938 & 1535 \\ \hline      
        $\Lambda$ & 1107 & 1820 \\ \hline
        $\Sigma$ & 1188 & 1897 \\ \hline       
        $\Xi$ & 1307 & 2172 \\ \hline        
        $\Delta$ & 1232 & 1829 \\ \hline 
\end{tabular}
\caption{List of vacuum masses for baryons included in the PD-CMF model. States differing only by isospin have the same mass in this model.}\label{tab:masses}
\end{table}

\bibliographystyle{JHEP}
\bibliography{reflist}

\providecommand{\href}[2]{#2}\begingroup\raggedright\begin{thebibliography}{10}

\bibitem{Tan:2021ahl}
H.~Tan, T.~Dore, V.~Dexheimer, J.~Noronha-Hostler and N.~Yunes, \emph{{Extreme matter meets extreme gravity: Ultraheavy neutron stars with phase transitions}}, \href{http://dx.doi.org/10.1103/PhysRevD.105.023018}{\emph{Phys. Rev. D} {\bf 105} (2022) 023018}, [\href{https://arxiv.org/abs/2106.03890}{{\tt 2106.03890}}].

\bibitem{Tan:2021nat}
H.~Tan, V.~Dexheimer, J.~Noronha-Hostler and N.~Yunes, \emph{{Finding Structure in the Speed of Sound of Supranuclear Matter from Binary Love Relations}}, \href{http://dx.doi.org/10.1103/PhysRevLett.128.161101}{\emph{Phys. Rev. Lett.} {\bf 128} (2022) 161101}, [\href{https://arxiv.org/abs/2111.10260}{{\tt 2111.10260}}].

\bibitem{Alford:2004pf}
M.~Alford, M.~Braby, M.~W. Paris and S.~Reddy, \emph{{Hybrid stars that masquerade as neutron stars}}, \href{http://dx.doi.org/10.1086/430902}{\emph{Astrophys. J.} {\bf 629} (2005) 969--978}, [\href{https://arxiv.org/abs/nucl-th/0411016}{{\tt nucl-th/0411016}}].

\bibitem{Hammond:2025kki}
P.~Hammond et~al., \emph{{Investigating the Impact of Higher-Order Phase Transitions in Binary Neutron-Star Mergers}},  \href{https://arxiv.org/abs/2508.10698}{{\tt 2508.10698}}.

\bibitem{Potekhin:2020ttj}
A.~Y. Potekhin, D.~A. Zyuzin, D.~G. Yakovlev, M.~V. Beznogov and Y.~A. Shibanov, \emph{{Thermal luminosities of cooling neutron stars}}, \href{http://dx.doi.org/10.1093/mnras/staa1871}{\emph{Mon. Not. Roy. Astron. Soc.} {\bf 496} (2020) 5052--5071}, [\href{https://arxiv.org/abs/2006.15004}{{\tt 2006.15004}}].

\bibitem{Dexheimer:2012eu}
V.~Dexheimer, J.~Steinheimer, R.~Negreiros and S.~Schramm, \emph{{Hybrid Stars in an SU(3) parity doublet model}}, \href{http://dx.doi.org/10.1103/PhysRevC.87.015804}{\emph{Phys. Rev. C} {\bf 87} (2013) 015804}, [\href{https://arxiv.org/abs/1206.3086}{{\tt 1206.3086}}].

\bibitem{Mukherjee:2017jzi}
A.~Mukherjee, S.~Schramm, J.~Steinheimer and V.~Dexheimer, \emph{{The application of the Quark-Hadron Chiral Parity-Doublet Model to neutron star matter}}, \href{http://dx.doi.org/10.1051/0004-6361/201731505}{\emph{Astron. Astrophys.} {\bf 608} (2017) A110}, [\href{https://arxiv.org/abs/1706.09191}{{\tt 1706.09191}}].

\bibitem{Marczenko:2018jui}
M.~Marczenko, D.~Blaschke, K.~Redlich and C.~Sasaki, \emph{{Chiral symmetry restoration by parity doubling and the structure of neutron stars}}, \href{http://dx.doi.org/10.1103/PhysRevD.98.103021}{\emph{Phys. Rev. D} {\bf 98} (2018) 103021}, [\href{https://arxiv.org/abs/1805.06886}{{\tt 1805.06886}}].

\bibitem{Brodie:2025nww}
L.~Brodie and R.~D. Pisarski, \emph{{Parity-Doubled Nucleons Can Rapidly Cool Neutron Stars}}, \href{http://dx.doi.org/10.1103/bf66-1hr5}{\emph{Phys. Rev. Lett.} {\bf 135} (2025) 152702}, [\href{https://arxiv.org/abs/2501.02055}{{\tt 2501.02055}}].

\bibitem{Page:2004fy}
D.~Page, J.~M. Lattimer, M.~Prakash and A.~W. Steiner, \emph{{Minimal cooling of neutron stars: A New paradigm}}, \href{http://dx.doi.org/10.1086/424844}{\emph{Astrophys. J. Suppl.} {\bf 155} (2004) 623--650}, [\href{https://arxiv.org/abs/astro-ph/0403657}{{\tt astro-ph/0403657}}].

\bibitem{Page:2006ud}
D.~Page and S.~Reddy, \emph{{Dense Matter in Compact Stars: Theoretical Developments and Observational Constraints}}, \href{http://dx.doi.org/10.1146/annurev.nucl.56.080805.140600}{\emph{Ann. Rev. Nucl. Part. Sci.} {\bf 56} (2006) 327--374}, [\href{https://arxiv.org/abs/astro-ph/0608360}{{\tt astro-ph/0608360}}].

\bibitem{Negreiros:2010tf}
R.~Negreiros, V.~A. Dexheimer and S.~Schramm, \emph{{Quark core impact on hybrid star cooling}}, \href{http://dx.doi.org/10.1103/PhysRevC.85.035805}{\emph{Phys. Rev. C} {\bf 85} (2012) 035805}, [\href{https://arxiv.org/abs/1011.2233}{{\tt 1011.2233}}].

\bibitem{Negreiros:2011ak}
R.~Negreiros, S.~Schramm and F.~Weber, \emph{{Impact of Rotation-Driven Particle Repopulation on the Thermal Evolution of Pulsars}}, \href{http://dx.doi.org/10.1016/j.physletb.2012.12.046}{\emph{Phys. Lett. B} {\bf 718} (2013) 1176--1180}, [\href{https://arxiv.org/abs/1103.3870}{{\tt 1103.3870}}].

\bibitem{Negreiros:2012aw}
R.~Negreiros, S.~Schramm and F.~Weber, \emph{{Thermal Evolution of Neutron Stars in 2 Dimensions}}, \href{http://dx.doi.org/10.1103/PhysRevD.85.104019}{\emph{Phys. Rev. D} {\bf 85} (2012) 104019}, [\href{https://arxiv.org/abs/1201.2381}{{\tt 1201.2381}}].

\bibitem{Aarts:2017rrl}
G.~Aarts, C.~Allton, D.~De~Boni, S.~Hands, B.~J{\"a}ger, C.~Praki et~al., \emph{{Light baryons below and above the deconfinement transition: medium effects and parity doubling}}, \href{http://dx.doi.org/10.1007/JHEP06(2017)034}{\emph{JHEP} {\bf 06} (2017) 034}, [\href{https://arxiv.org/abs/1703.09246}{{\tt 1703.09246}}].

\bibitem{Detar:1988kn}
C.~E. Detar and T.~Kunihiro, \emph{{Linear $\sigma$ Model With Parity Doubling}}, \href{http://dx.doi.org/10.1103/PhysRevD.39.2805}{\emph{Phys. Rev. D} {\bf 39} (1989) 2805}.

\bibitem{Zschiesche:2006zj}
D.~Zschiesche, L.~Tolos, J.~Schaffner-Bielich and R.~D. Pisarski, \emph{{Cold, dense nuclear matter in a SU(2) parity doublet model}}, \href{http://dx.doi.org/10.1103/PhysRevC.75.055202}{\emph{Phys. Rev. C} {\bf 75} (2007) 055202}, [\href{https://arxiv.org/abs/nucl-th/0608044}{{\tt nucl-th/0608044}}].

\bibitem{Steinheimer:2011ea}
J.~Steinheimer, S.~Schramm and H.~Stocker, \emph{{The hadronic SU(3) Parity Doublet Model for Dense Matter, its extension to quarks and the strange equation of state}}, \href{http://dx.doi.org/10.1103/PhysRevC.84.045208}{\emph{Phys. Rev. C} {\bf 84} (2011) 045208}, [\href{https://arxiv.org/abs/1108.2596}{{\tt 1108.2596}}].

\bibitem{Sasaki:2017glk}
C.~Sasaki, \emph{{Parity doubling of baryons in a chiral approach with three flavors}}, \href{http://dx.doi.org/10.1016/j.nuclphysa.2018.01.004}{\emph{Nucl. Phys. A} {\bf 970} (2018) 388--397}, [\href{https://arxiv.org/abs/1707.05081}{{\tt 1707.05081}}].

\bibitem{Minamikawa:2020jfj}
T.~Minamikawa, T.~Kojo and M.~Harada, \emph{{Quark-hadron crossover equations of state for neutron stars: constraining the chiral invariant mass in a parity doublet model}}, \href{http://dx.doi.org/10.1103/PhysRevC.103.045205}{\emph{Phys. Rev. C} {\bf 103} (2021) 045205}, [\href{https://arxiv.org/abs/2011.13684}{{\tt 2011.13684}}].

\bibitem{Marczenko:2022hyt}
M.~Marczenko, K.~Redlich and C.~Sasaki, \emph{{Chiral symmetry restoration and {\ensuremath{\Delta}} matter formation in neutron stars}}, \href{http://dx.doi.org/10.1103/PhysRevD.105.103009}{\emph{Phys. Rev. D} {\bf 105} (2022) 103009}, [\href{https://arxiv.org/abs/2203.00269}{{\tt 2203.00269}}].

\bibitem{Fraga:2023wtd}
E.~S. Fraga, R.~da~Mata and J.~Schaffner-Bielich, \emph{{SU(3) parity doubling in cold neutron star matter}}, \href{http://dx.doi.org/10.1103/PhysRevD.108.116003}{\emph{Phys. Rev. D} {\bf 108} (2023) 116003}, [\href{https://arxiv.org/abs/2309.02368}{{\tt 2309.02368}}].

\bibitem{Eser:2024xil}
J.~Eser and J.-P. Blaizot, \emph{{Thermodynamics of the parity-doublet model. II. Asymmetric and neutron matter}}, \href{http://dx.doi.org/10.1103/PhysRevC.110.065205}{\emph{Phys. Rev. C} {\bf 110} (2024) 065205}, [\href{https://arxiv.org/abs/2408.01302}{{\tt 2408.01302}}].

\bibitem{Gao:2026scv}
B.~Gao, \emph{{Chiral symmetry restoration and hyperon suppression in neutron stars}},  \href{https://arxiv.org/abs/2602.12503}{{\tt 2602.12503}}.

\bibitem{Ratti:2005jh}
C.~Ratti, M.~A. Thaler and W.~Weise, \emph{{Phases of QCD: Lattice thermodynamics and a field theoretical model}}, \href{http://dx.doi.org/10.1103/PhysRevD.73.014019}{\emph{Phys. Rev. D} {\bf 73} (2006) 014019}, [\href{https://arxiv.org/abs/hep-ph/0506234}{{\tt hep-ph/0506234}}].

\bibitem{Roessner:2006xn}
S.~Roessner, C.~Ratti and W.~Weise, \emph{{Polyakov loop, diquarks and the two-flavour phase diagram}}, \href{http://dx.doi.org/10.1103/PhysRevD.75.034007}{\emph{Phys. Rev. D} {\bf 75} (2007) 034007}, [\href{https://arxiv.org/abs/hep-ph/0609281}{{\tt hep-ph/0609281}}].

\bibitem{Dexheimer:2009hi}
V.~A. Dexheimer and S.~Schramm, \emph{{A Novel Approach to Model Hybrid Stars}}, \href{http://dx.doi.org/10.1103/PhysRevC.81.045201}{\emph{Phys. Rev. C} {\bf 81} (2010) 045201}, [\href{https://arxiv.org/abs/0901.1748}{{\tt 0901.1748}}].

\bibitem{Papazoglou:1998vr}
P.~Papazoglou, D.~Zschiesche, S.~Schramm, J.~Schaffner-Bielich, H.~Stoecker and W.~Greiner, \emph{{Nuclei in a chiral SU(3) model}}, \href{http://dx.doi.org/10.1103/PhysRevC.59.411}{\emph{Phys. Rev. C} {\bf 59} (1999) 411--427}, [\href{https://arxiv.org/abs/nucl-th/9806087}{{\tt nucl-th/9806087}}].

\bibitem{Dexheimer:2008ax}
V.~Dexheimer and S.~Schramm, \emph{{Proto-Neutron and Neutron Stars in a Chiral SU(3) Model}}, \href{http://dx.doi.org/10.1086/589735}{\emph{Astrophys. J.} {\bf 683} (2008) 943--948}, [\href{https://arxiv.org/abs/0802.1999}{{\tt 0802.1999}}].

\bibitem{Motornenko:2020yme}
A.~Motornenko, S.~Pal, A.~Bhattacharyya, J.~Steinheimer and H.~Stoecker, \emph{{Repulsive properties of hadrons in lattice QCD data and neutron stars}}, \href{http://dx.doi.org/10.1103/PhysRevC.103.054908}{\emph{Phys. Rev. C} {\bf 103} (2021) 054908}, [\href{https://arxiv.org/abs/2009.10848}{{\tt 2009.10848}}].

\bibitem{Steinheimer:2025hsr}
J.~Steinheimer, M.~Omana~Kuttan, T.~Reichert, Y.~Nara and M.~Bleicher, \emph{{Simultaneous description of high density QCD matter in heavy ion collisions and neutron star observations}}, \href{http://dx.doi.org/10.1016/j.physletb.2025.139605}{\emph{Phys. Lett. B} {\bf 867} (2025) 139605}, [\href{https://arxiv.org/abs/2501.12849}{{\tt 2501.12849}}].

\bibitem{ParticleDataGroup:2022pth}
{\scshape Particle Data Group} collaboration, R.~L. Workman et~al., \emph{{Review of Particle Physics}}, \href{http://dx.doi.org/10.1093/ptep/ptac097}{\emph{PTEP} {\bf 2022} (2022) 083C01}.

\bibitem{PhysRevD.108.014510}
{\scshape HotQCD Collaboration} collaboration, D.~Bollweg, D.~A. Clarke, J.~Goswami, O.~Kaczmarek, F.~Karsch, S.~Mukherjee et~al., \emph{Equation of state and speed of sound of ($2+1$)-flavor qcd in strangeness-neutral matter at nonvanishing net baryon-number density}, \href{http://dx.doi.org/10.1103/PhysRevD.108.014510}{\emph{Phys. Rev. D} {\bf 108} (Jul, 2023) 014510}.

\bibitem{Borsanyi:2012cr}
S.~Borsanyi, G.~Endrodi, Z.~Fodor, S.~D. Katz, S.~Krieg, C.~Ratti et~al., \emph{{QCD equation of state at nonzero chemical potential: continuum results with physical quark masses at order $mu^2$}}, \href{http://dx.doi.org/10.1007/JHEP08(2012)053}{\emph{JHEP} {\bf 08} (2012) 053}, [\href{https://arxiv.org/abs/1204.6710}{{\tt 1204.6710}}].

\bibitem{Minamikawa:2021fln}
T.~Minamikawa, T.~Kojo and M.~Harada, \emph{{Chiral condensates for neutron stars in hadron-quark crossover: From a parity doublet nucleon model to a Nambu{\textendash}Jona-Lasinio quark model}}, \href{http://dx.doi.org/10.1103/PhysRevC.104.065201}{\emph{Phys. Rev. C} {\bf 104} (2021) 065201}, [\href{https://arxiv.org/abs/2107.14545}{{\tt 2107.14545}}].

\bibitem{Minamikawa:2022ckn}
T.~Minamikawa, T.~Kojo and M.~Harada, \emph{{Chiral condensates for neutron stars in hadron-quark crossover; from a parity doublet nucleon model to an NJL quark model}}, \href{http://dx.doi.org/10.31349/SuplRevMexFis.3.0308121}{\emph{Rev. Mex. Fis. Suppl.} {\bf 3} (2022) 0308121}, [\href{https://arxiv.org/abs/2202.04873}{{\tt 2202.04873}}].

\bibitem{Gao:2022klm}
B.~Gao, T.~Minamikawa, T.~Kojo and M.~Harada, \emph{{Impacts of the U(1)A anomaly on nuclear and neutron star equation~of state based on a parity doublet model}}, \href{http://dx.doi.org/10.1103/PhysRevC.106.065205}{\emph{Phys. Rev. C} {\bf 106} (2022) 065205}, [\href{https://arxiv.org/abs/2207.05970}{{\tt 2207.05970}}].

\bibitem{Minamikawa:2023eky}
T.~Minamikawa, B.~Gao, T.~Kojo and M.~Harada, \emph{{Chiral Restoration of Nucleons in Neutron Star Matter: Studies Based on a Parity Doublet Model}}, \href{http://dx.doi.org/10.3390/sym15030745}{\emph{Symmetry} {\bf 15} (2023) 745}, [\href{https://arxiv.org/abs/2302.00825}{{\tt 2302.00825}}].

\bibitem{Minamikawa:2023ypn}
T.~Minamikawa, B.~Gao, T.~kojo and M.~Harada, \emph{{Parity doublet model for baryon octets: Diquark classifications and mass hierarchy based on the quark-line diagram}}, \href{http://dx.doi.org/10.1103/PhysRevD.108.076017}{\emph{Phys. Rev. D} {\bf 108} (2023) 076017}, [\href{https://arxiv.org/abs/2306.15564}{{\tt 2306.15564}}].

\bibitem{Gao:2024mew}
B.~Gao, T.~Kojo and M.~Harada, \emph{{Parity doublet model for baryon octets: Ground states saturated by good diquarks and the role of bad diquarks for excited states}}, \href{http://dx.doi.org/10.1103/PhysRevD.110.016016}{\emph{Phys. Rev. D} {\bf 110} (2024) 016016}, [\href{https://arxiv.org/abs/2403.18214}{{\tt 2403.18214}}].

\bibitem{Fukushima:2003fw}
K.~Fukushima, \emph{{Chiral effective model with the Polyakov loop}}, \href{http://dx.doi.org/10.1016/j.physletb.2004.04.027}{\emph{Phys. Lett. B} {\bf 591} (2004) 277--284}, [\href{https://arxiv.org/abs/hep-ph/0310121}{{\tt hep-ph/0310121}}].

\bibitem{Motornenko:2019arp}
A.~Motornenko, J.~Steinheimer, V.~Vovchenko, S.~Schramm and H.~Stoecker, \emph{{Equation of state for hot QCD and compact stars from a mean field approach}}, \href{http://dx.doi.org/10.1103/PhysRevC.101.034904}{\emph{Phys. Rev. C} {\bf 101} (2020) 034904}, [\href{https://arxiv.org/abs/1905.00866}{{\tt 1905.00866}}].

\bibitem{Rischke:1991ke}
D.~H. Rischke, M.~I. Gorenstein, H.~Stoecker and W.~Greiner, \emph{{Excluded volume effect for the nuclear matter equation of state}}, \href{http://dx.doi.org/10.1007/BF01548574}{\emph{Z. Phys. C} {\bf 51} (1991) 485--490}.

\bibitem{Baym:1971pw}
G.~Baym, C.~Pethick and P.~Sutherland, \emph{{The Ground state of matter at high densities: Equation of state and stellar models}}, \href{http://dx.doi.org/10.1086/151216}{\emph{Astrophys. J.} {\bf 170} (1971) 299--317}.

\bibitem{Tolman}
R.~C. Tolman, \emph{Static solutions of einstein's field equations for spheres of fluid}, \href{http://dx.doi.org/10.1103/PhysRev.55.364}{\emph{Phys. Rev.} {\bf 55} (Feb, 1939) 364--373}.

\bibitem{OppenheimerVolkoff}
J.~R. Oppenheimer and G.~M. Volkoff, \emph{On massive neutron cores}, \href{http://dx.doi.org/10.1103/PhysRev.55.374}{\emph{Phys. Rev.} {\bf 55} (Feb, 1939) 374--381}.

\bibitem{2006NuPhA.777..497P}
D.~Page, U.~Geppert and F.~Weber, \emph{{The cooling of compact stars}}, \href{http://dx.doi.org/10.1016/j.nuclphysa.2005.09.019}{\emph{Nuclear Physics A} {\bf 777} (2006) 497--530}.

\bibitem{1999Weber..book}
F.~Weber, \emph{{Pulsars as astrophysical laboratories for nuclear and particle physics}}.
\newblock Institute of Physics, Bristol, U.K., 1999.

\bibitem{1996NuPhA.605..531S}
C.~Schaab, F.~Weber, M.~K. Weigel and N.~K. Glendenning, \emph{{Thermal evolution of compact stars}}, \href{http://dx.doi.org/10.1016/0375-9474(96)00164-9}{\emph{Nuclear Physics A} {\bf 605} (aug, 1996) 531--565}.

\bibitem{Yakovlev2000}
D.~G. Yakovlev, A.~D. Kaminker, O.~Y. Gnedin and P.~Haensel, \emph{{Neutrino Emission from Neutron Stars}}, \href{http://dx.doi.org/10.1016/S0370-1573(00)00131-9}{\emph{Physics Reports} {\bf 354} (2000) 1--155}, [\href{https://arxiv.org/abs/0012122}{{\tt 0012122}}].

\bibitem{Yakovlev2004}
D.~Yakovlev and C.~Pethick, \emph{{NEUTRON STAR COOLING}}, \href{http://dx.doi.org/10.1146/annurev.astro.42.053102.134013}{\emph{Annual Review of Astronomy and Astrophysics} {\bf 42} (sep, 2004) 169--210}, [\href{https://arxiv.org/abs/0402143}{{\tt 0402143}}].

\bibitem{gnedin1995thermal}
O.~Y. Gnedin and D.~Yakovlev, \emph{Thermal conductivity of electrons and muons in neutron star cores}, {\emph{Nuclear Physics A} {\bf 582} (1995) 697--716}.

\bibitem{potekhin2015neutron}
A.~Y. Potekhin, J.~A. Pons and D.~Page, \emph{Neutron stars—cooling and transport}, {\emph{Space Science Reviews} {\bf 191} (2015) 239--291}.

\bibitem{Friman1979NeutrinoEmission}
B.~L. Friman and O.~V. Maxwell, \emph{Neutrino emissivities of neutron stars}, \href{http://dx.doi.org/10.1086/157313}{\emph{The Astrophysical Journal} {\bf 232} (1979) 541--557}.

\bibitem{Lattimer1991DirectUrca}
J.~M. Lattimer, C.~J. Pethick, M.~Prakash and P.~Haensel, \emph{Direct urca process in neutron stars}, \href{http://dx.doi.org/10.1103/PhysRevLett.66.2701}{\emph{Physical Review Letters} {\bf 66} (1991) 2701--2704}.

\bibitem{Itoh1996NeutrinoProcesses}
N.~Itoh, H.~Hayashi, A.~Nishikawa and Y.~Kohyama, \emph{Neutrino energy loss in stellar interiors. vii. pair, photo-, plasma, and bremsstrahlung processes}, \href{http://dx.doi.org/10.1086/192264}{\emph{The Astrophysical Journal Supplement Series} {\bf 102} (1996) 411--424}.

\bibitem{Yakovlev2001NeutronStarCooling}
D.~G. Yakovlev, A.~D. Kaminker, O.~Y. Gnedin and P.~Haensel, \emph{Neutrino emission from neutron stars}, \href{http://dx.doi.org/10.1016/S0370-1573(00)00131-9}{\emph{Physics Reports} {\bf 354} (2001) 1--155}.

\bibitem{Iwamoto1982QuarkCooling}
N.~Iwamoto, \emph{Neutrino emissivities and mean free paths of degenerate quark matter}, \href{http://dx.doi.org/10.1016/0003-4916(82)90206-0}{\emph{Annals of Physics} {\bf 141} (1982) 1--49}.

\bibitem{Yakovlev:2000jp}
D.~G. Yakovlev, A.~D. Kaminker, O.~Y. Gnedin and P.~Haensel, \emph{{Neutrino emission from neutron stars}}, \href{http://dx.doi.org/10.1016/S0370-1573(00)00131-9}{\emph{Phys. Rept.} {\bf 354} (2001) 1}, [\href{https://arxiv.org/abs/astro-ph/0012122}{{\tt astro-ph/0012122}}].

\bibitem{Lattimer:1991ib}
J.~M. Lattimer, M.~Prakash, C.~J. Pethick and P.~Haensel, \emph{{Direct URCA process in neutron stars}}, \href{http://dx.doi.org/10.1103/PhysRevLett.66.2701}{\emph{Phys. Rev. Lett.} {\bf 66} (1991) 2701--2704}.

\bibitem{Friman:1979ecl}
B.~L. Friman and O.~V. Maxwell, \emph{{Neutron Star Neutrino Emissivities}}, \href{http://dx.doi.org/10.1086/157313}{\emph{Astrophys. J.} {\bf 232} (1979) 541--557}.

\bibitem{ho2015tests}
W.~C. Ho, K.~G. Elshamouty, C.~O. Heinke and A.~Y. Potekhin, \emph{Tests of the nuclear equation of state and superfluid and superconducting gaps using the cassiopeia a neutron star}, {\emph{Physical Review C} {\bf 91} (2015) 015806}.

\bibitem{chen1993pairing}
J.~Chen, J.~Clark, R.~Dav{\'e} and V.~Khodel, \emph{Pairing gaps in nucleonic superfluids}, {\emph{Nuclear Physics A} {\bf 555} (1993) 59--89}.

\bibitem{amundsen1985superfluidity}
L.~Amundsen and E.~{\O}stgaard, \emph{Superfluidity of neutron matter:(ii). triplet pairing}, {\emph{Nuclear Physics A} {\bf 442} (1985) 163--188}.

\bibitem{Sedrakian:2018ydt}
A.~Sedrakian and J.~W. Clark, \emph{{Superfluidity in nuclear systems and neutron stars}}, \href{http://dx.doi.org/10.1140/epja/i2019-12863-6}{\emph{Eur. Phys. J. A} {\bf 55} (2019) 167}, [\href{https://arxiv.org/abs/1802.00017}{{\tt 1802.00017}}].

\bibitem{Page:2013hxa}
D.~Page, J.~M. Lattimer, M.~Prakash and A.~W. Steiner, \emph{Stellar superfluids},  Feburary, 2013.

\bibitem{Anzuini:2021rjv}
F.~Anzuini, A.~Melatos, C.~Dehman, D.~Vigan{\`o} and J.~A. Pons, \emph{{Fast cooling and internal heating in hyperon stars}}, \href{http://dx.doi.org/10.1093/mnras/stab3126}{\emph{Mon. Not. Roy. Astron. Soc.} {\bf 509} (2021) 2609--2623}, [\href{https://arxiv.org/abs/2110.14039}{{\tt 2110.14039}}].

\bibitem{raduta2018cooling}
A.~R. Raduta, A.~Sedrakian and F.~Weber, \emph{Cooling of hypernuclear compact stars}, {\emph{Monthly Notices of the Royal Astronomical Society} {\bf 475} (2018) 4347--4356}.

\bibitem{PhysRevD.103.083004}
M.~Fortin, A.~R. Raduta, S.~Avancini and C.~m.~c. Provid\^encia, \emph{Thermal evolution of relativistic hyperonic compact stars with calibrated equations of state}, \href{http://dx.doi.org/10.1103/PhysRevD.103.083004}{\emph{Phys. Rev. D} {\bf 103} (Apr, 2021) 083004}.

\bibitem{Raduta:2019rsk}
A.~R. Raduta, J.~J. Li, A.~Sedrakian and F.~Weber, \emph{{Cooling of hypernuclear compact stars: Hartree{\textendash}Fock models and high-density pairing}}, \href{http://dx.doi.org/10.1093/mnras/stz1459}{\emph{Mon. Not. Roy. Astron. Soc.} {\bf 487} (2019) 2639--2652}, [\href{https://arxiv.org/abs/1903.01295}{{\tt 1903.01295}}].

\bibitem{Grigorian:2018bvg}
H.~Grigorian, D.~N. Voskresensky and K.~A. Maslov, \emph{{Cooling of neutron stars in {\textquotedblleft}nuclear medium cooling scenario{\textquotedblright} with stiff equation of state including hyperons}}, \href{http://dx.doi.org/10.1016/j.nuclphysa.2018.10.014}{\emph{Nucl. Phys. A} {\bf 980} (2018) 105--130}, [\href{https://arxiv.org/abs/1808.01819}{{\tt 1808.01819}}].

\bibitem{Fortin:2021umb}
M.~Fortin, A.~R. Raduta, S.~Avancini and C.~Provid{\^e}ncia, \emph{{Thermal evolution of relativistic hyperonic compact stars with calibrated equations of state}}, \href{http://dx.doi.org/10.1103/PhysRevD.103.083004}{\emph{Phys. Rev. D} {\bf 103} (2021) 083004}, [\href{https://arxiv.org/abs/2102.07565}{{\tt 2102.07565}}].

\bibitem{Raduta:2017wpp}
A.~R. Raduta, A.~Sedrakian and F.~Weber, \emph{{Cooling of hypernuclear compact stars}}, \href{http://dx.doi.org/10.1093/mnras/stx3318}{\emph{Mon. Not. Roy. Astron. Soc.} {\bf 475} (2018) 4347--4356}, [\href{https://arxiv.org/abs/1712.00584}{{\tt 1712.00584}}].

\bibitem{IWAMOTO19821}
N.~Iwamoto, \emph{Neutrino emissivities and mean free paths of degenerate quark matter}, \href{http://dx.doi.org/https://doi.org/10.1016/0003-4916(82)90271-8}{\emph{Annals of Physics} {\bf 141} (1982) 1--49}.

\bibitem{PhysRevLett.44.1637}
N.~Iwamoto, \emph{Quark beta decay and the cooling of neutron stars}, \href{http://dx.doi.org/10.1103/PhysRevLett.44.1637}{\emph{Phys. Rev. Lett.} {\bf 44} (Jun, 1980) 1637--1640}.

\bibitem{PhysRevD.66.063003}
P.~Jaikumar, M.~Prakash and T.~Sch\"afer, \emph{Neutrino emission from goldstone modes in dense quark matter}, \href{http://dx.doi.org/10.1103/PhysRevD.66.063003}{\emph{Phys. Rev. D} {\bf 66} (Sep, 2002) 063003}.

\bibitem{Alford:1998mk}
M.~G. Alford, K.~Rajagopal and F.~Wilczek, \emph{{Color flavor locking and chiral symmetry breaking in high density QCD}}, \href{http://dx.doi.org/10.1016/S0550-3213(98)00668-3}{\emph{Nucl. Phys. B} {\bf 537} (1999) 443--458}, [\href{https://arxiv.org/abs/hep-ph/9804403}{{\tt hep-ph/9804403}}].

\bibitem{Alford:2007xm}
M.~G. Alford, A.~Schmitt, K.~Rajagopal and T.~Sch{\"a}fer, \emph{{Color superconductivity in dense quark matter}}, \href{http://dx.doi.org/10.1103/RevModPhys.80.1455}{\emph{Rev. Mod. Phys.} {\bf 80} (2008) 1455--1515}, [\href{https://arxiv.org/abs/0709.4635}{{\tt 0709.4635}}].

\bibitem{Rapp:1997zu}
R.~Rapp, T.~Sch{\"a}fer, E.~V. Shuryak and M.~Velkovsky, \emph{{Diquark Bose condensates in high density matter and instantons}}, \href{http://dx.doi.org/10.1103/PhysRevLett.81.53}{\emph{Phys. Rev. Lett.} {\bf 81} (1998) 53--56}, [\href{https://arxiv.org/abs/hep-ph/9711396}{{\tt hep-ph/9711396}}].

\bibitem{grigorian2005cooling}
H.~Grigorian, D.~Blaschke and D.~Voskresensky, \emph{Cooling of neutron stars with color superconducting quark cores}, {\emph{Physical Review C—Nuclear Physics} {\bf 71} (2005) 045801}.

\bibitem{Blaschke2001}
D.~Blaschke, \emph{Cooling of hybrid neutron stars and hypothetical self-bound objects with superconducting quark cores}, \href{http://dx.doi.org/10.1051/0004-6361}{\emph{Astronomy and Astrophysics} {\bf 568} (2001) 561--568}.

\bibitem{Gudmundsson1982}
E.~Gudmundsson, \emph{Neutron star envelopes}, {\emph{The Astrophysical Journal} {\bf 259} (1982) L19}.

\bibitem{Potekhin_2003}
A.~Y. Potekhin, D.~G. Yakovlev, G.~Chabrier and O.~Y. Gnedin, \emph{Thermal structure and cooling of superfluid neutron stars with accreted magnetized envelopes}, \href{http://dx.doi.org/10.1086/376900}{\emph{The Astrophysical Journal} {\bf 594} (sep, 2003) 404}.

\bibitem{Beznogov:2014yia}
M.~V. Beznogov and D.~G. Yakovlev, \emph{{Statistical theory of thermal evolution of neutron stars}}, \href{http://dx.doi.org/10.1093/mnras/stu2506}{\emph{Mon. Not. Roy. Astron. Soc.} {\bf 447} (2015) 1598--1609}, [\href{https://arxiv.org/abs/1411.6803}{{\tt 1411.6803}}].

\bibitem{Doroshenko:2022}
V.~Doroshenko, V.~Suleimanov, G.~Pühlhofer and A.~Santangelo, \emph{{A strangely light neutron star within a supernova remnant}}, \href{http://dx.doi.org/10.1038/s41550-022-01800-1}{\emph{Nature Astronomy} (2022) }.

\bibitem{Gudmundsson1983}
E.~H. Gudmundsson, C.~J. Pethick and R.~I. Epstein, \emph{Structure of neutron star envelopes}, \href{http://dx.doi.org/10.1086/161292}{\emph{The Astrophysical Journal} {\bf 272} (9, 1983) 286}.

\bibitem{Potekhin1997}
A.~Y. Potekhin, G.~Chabrier and D.~G. Yakovlev, \emph{Internal temperatures and cooling of neutron stars with accreted envelopes}, {\emph{Astronomy \& Astrophysics} {\bf 323} (6, 1997) 415}.

\bibitem{Zhang_2025}
S.~R. Zhang, J.~A. Rueda~Hernandez and R.~Negreiros, \emph{Can the central compact object in hess j1731–347 be indeed the lightest neutron star observed?}, \href{http://dx.doi.org/10.3847/1538-4357/ad96b5}{\emph{The Astrophysical Journal} {\bf 978} (dec, 2024) 1}.

\bibitem{sagun2023nature}
V.~Sagun, E.~Giangrandi, T.~Dietrich, O.~Ivanytskyi, R.~Negreiros and C.~Provid{\^e}ncia, \emph{What is the nature of the hess j1731-347 compact object?}, {\emph{The Astrophysical Journal} {\bf 958} (2023) 49}.

\bibitem{Heide:1993yz}
E.~K. Heide, S.~Rudaz and P.~J. Ellis, \emph{{An Effective Lagrangian with broken scale and chiral symmetry applied to nuclear matter and finite nuclei}}, \href{http://dx.doi.org/10.1016/0375-9474(94)90717-X}{\emph{Nucl. Phys. A} {\bf 571} (1994) 713--732}, [\href{https://arxiv.org/abs/nucl-th/9308002}{{\tt nucl-th/9308002}}].

\bibitem{Papazoglou:1996hf}
P.~Papazoglou, J.~Schaffner, S.~Schramm, D.~Zschiesche, H.~Stoecker and W.~Greiner, \emph{{Phase transition in the chiral sigma - omega model with dilatons}}, \href{http://dx.doi.org/10.1103/PhysRevC.55.1499}{\emph{Phys. Rev. C} {\bf 55} (1997) 1499--1508}, [\href{https://arxiv.org/abs/nucl-th/9609035}{{\tt nucl-th/9609035}}].

\bibitem{Dexheimer:2015qha}
V.~Dexheimer, R.~Negreiros and S.~Schramm, \emph{{Reconciling Nuclear and Astrophysical Constraints}}, \href{http://dx.doi.org/10.1103/PhysRevC.92.012801}{\emph{Phys. Rev. C} {\bf 92} (2015) 012801}, [\href{https://arxiv.org/abs/1503.07785}{{\tt 1503.07785}}].

\bibitem{Dexheimer:2018dhb}
V.~Dexheimer, R.~de~Oliveira~Gomes, S.~Schramm and H.~Pais, \emph{{What do we learn about vector interactions from GW170817?}}, \href{http://dx.doi.org/10.1088/1361-6471/ab01f0}{\emph{J. Phys. G} {\bf 46} (2019) 034002}, [\href{https://arxiv.org/abs/1810.06109}{{\tt 1810.06109}}].

\bibitem{Kumar:2024owe}
R.~Kumar, Y.~Wang, N.~C. Camacho, A.~Kumar, J.~Noronha-Hostler and V.~Dexheimer, \emph{{Modern nuclear and astrophysical constraints of dense matter in a redefined chiral approach}}, \href{http://dx.doi.org/10.1103/PhysRevD.109.074008}{\emph{Phys. Rev. D} {\bf 109} (2024) 074008}, [\href{https://arxiv.org/abs/2401.12944}{{\tt 2401.12944}}].

\end{thebibliography}\endgroup

\end{document}